*Title:*

*A Twist-Bend Chiral Helix of 8nm Pitch*

*in a Nematic Liquid Crystal of Achiral Molecular Dimers*

*Authors:*

Dong Chen[1], Jan H. Porada[2], Justin B. Hooper[3], Arthur Klittnick[1], Yongqiang Shen[1]

Eva Korblova[2], Dmitry Bedrov[3], David M. Walba[2],

Matthew A. Glaser[1], Joseph E. Maclennan[1] and Noel A. Clark[1]

*Affiliations:*

[1] Department of Physics and Liquid Crystal Materials Research Center, University of Colorado, Boulder, CO 80309-0390, USA.

[2] Department of Chemistry and Biochemistry and Liquid Crystal Materials Research Center, University of Colorado, Boulder, CO 80309-0215, USA.

[3] Department of Materials Science and Engineering, University of Utah, Salt Lake City, UT 84112, and Liquid Crystal Materials Research Center, University of Colorado, Boulder, CO 80309-0215, USA.

*Abstract:*

Freeze Fracture Transmission Electron Microscopy (FFTEM) study of the nanoscale structure of the so-called "twist-bend" nematic (NX) phase of the cyanobiphenyl (CB) dimer molecule $CB(CH_2)_7CB$ reveals a stripe texture of fluid layers periodically arrayed with a bulk spacing of d ≈ 8.3 nm. Fluidity and a rigorously maintained spacing produce long-range-ordered fluid layered focal conic domains. Absence of a lamellar x-ray reflection at wavevector $q \approx 2\pi/8$ nm$^{-1}$ or its harmonics in synchrotron-based scattering experiments indicates that this periodic structure is achieved with no detectable associated modulation of the electron density, and thus has nematic molecular ordering. A search for periodic ordering with d ~ 8nm in $CB(CH_2)_7CB$ using atomistic molecular dynamic computer simulation yielded equilibration of a conical twist-bend helixed nematic ground state, of the sort first proposed by Meyer, and envisioned in systems of bent molecules by Dozov and Memmer, We identify θ ≈ 33º as the cone angle, and p ≈ 8nm as the full pitch of the helix, the shortest ever found in a nematic fluid.

*Main Text:*

Recently there has been great interest in the liquid crystal (LC) phase behavior of achiral dimer molecules, such as cyanobiphenyl-$(CH_2)_n$-cyanobiphenyl (CBnCB), shown for n = 7 in Figure 1A (*1,2*). This attention results from the observation of a transition in such mesogens from a typical nematic (N) to a lower temperature (NX) phase, also apparently nematic, that exhibits a variety of unusual characteristics (*3,4,5,6,7,8,9,10*). These include: (**i**) textural features in depolarized transmission light microscopy (DTLM) similar to those found in fluid lamellar smectic phases but with no x-ray scattering to indicate lamellar ordering of molecules (*8*); (**ii**) A variety of other, completely unfamiliar DTLM textures (*6*), including the spontaneous appearance of director field deformation and evidence for small Frank elastic constants (*3*); (**iii**) evidence for the chiral molecular organization on the NMR time scale (*4*), and in macroscopic conglomerate domains in EO experiments on monodomain textures (*9*); (**iv**) distinctive n-odd/n-even effects, including, in particular, that (**i**) – (**iii**) are found only in the n-odd homologs (*6*).

These observations, combined with the fact the all-trans conformations of the n-odd homologous dimers are distinctly bent (Figure 1B), has led to the notion that the NX is a "twist-bend" (TB) phase, sketched in Figure 1C, a nematic having a conically helixed ground state, of the sort originally proposed by Meyer as the result of the spontaneous appearance of bend flexoelectric polarization (*11*). More recently Dozov proposed such a ground state as a spontaneously chiral conglomerate domain stabilized by molecular bend (*12*), and Memmer obtained such structures in computer simulations of systems of bent Gay-Berne dimers (*13*). This ground state helix can be written for CB7CB in terms of a half-molecular director, **n**(z), in this case the local average orientation of the long axis of a biphenyl half-molecular core, taken to located at the center of the bond between its rings (Figure 1B). Given by **n**(z) = **z**cosθ + sinθ(**x**sinφ + **y**cosφ) (*11*), the half-molecular director of the TB ground state reorients on a cone of angle θ coaxial with the helix axis **z** through azimuthal angle φ = $2\pi z/p_{TB}$, where $p_{TB}$ is the pitch, in a fashion similar to that of the molecular director helix in a chiral smectic C, shown in Figure 1D, but in this case without smectic layering. Since such a helical nemat-ic has no molecular positional layering and, in addition, has the same signs and magnitudes of director deformation (bend and twist) everywhere, it has complete translational symmetry of director distortion magnitude and thus should not exhibit Bragg diffraction from the helix using scattering probes of either mass or electron density, such as nonresonant x-ray or neutron scattering (*14*). To date there have been neither structural observations nor confirmation of a nematic TB ground state, nor measurement of pitch, $p_{TB}$, its basic parameter, predicted by both modeling (*12*) and suggested in simulations of bent core mesogens (*13*) to be on the order of a few molecular lengths. Here we report the results of a combined Freeze-Fracture Transmission Electron Microscopy (FFTEM), synchrotron x-ray diffraction (XRD), and atomistic molecular dynamic (MD) computer simulation study of the nanoscale structure of the NX phase of CB7CB. The FFTEM and XRD reveal a periodic nanoscale structure of fluid layers that is of spatially uniform density, and thus nematic. Equilibration of structures of the measured p ~ 8nm layer spacing using MD simulation robustly relaxes to the helical conical nematic TB ground state of Meyer (*11*), Dozov (*12*) and Memmer (*13*). We will therefore follow Ref. (*4*) and refer to the NX phase of CB7CB as the NTB phase.

CB7CB (1) was synthesized from 1,7-bis(4'-bromophenyl)heptane-1,7-dione (*15*), and the resulting sample characterized with respect to its LC phase behavior using differential scanning calorimetry and Depolarized Transmitted Light Microscopy (DTLM) with a temperature controlled hot-stage. The observed transitions, isotropic (I) → 112ºC → N → 99ºC → NTB (cooling), and I ← 113ºC ← N ← 100ºC ← NTB (heating), are in substantial agreement with literature values (*1,2,8,16*). FFTEM experiments were carried out by filling a few micron thick gap between 4mm by 6mm glass plates with CB7CB, heating the cell to the isotropic phase and cooling under hot-stage DTLM observation into the N and NTB phases. The N phase showed typical Schlieren DTLM textures and the NTB phase appeared via a first-order phase transition as slightly lower birefringence domains of a broken fan texture. Once T was reduced and stabilized to a chosen temperature, the sample was rapidly quenched to T < 90K by ejection from the hot-stage into liquid propane, and then maintained at T = 77K in liquid nitrogen. As shown in x-ray scattering experiments (see Figure S16) the NTB phase readily supercools to room temperature even under conditions of slow cooling, in agreement with prior observations (16), so that there is no chance for any ordering but that of the N or NTB to appear in samples so rapidly quenched. Samples were transferred cold into a Balzers BAF-060 freeze fracture apparatus and the LC fractured by pulling apart the glass plates cold under vacuum. Shadowing of the topographic structure of the fracture face was carried out by oblique evaporation of a 1.5 nm thick platinum film, which was embedded in a thicker evaporated carbon film to form an electron absorption replica with which the interface topography could be viewed in a TEM.

Typical fracture topography of CB7CB in the NTB phase (T = 90 ºC) is shown in Figure 2, where the red arrow indicates the azimuthal direction of incidence of the Pt evaporation, and the image is such that the shadowed areas having less Pt are darkest. A pattern of quasi-periodic curved stripes, of spacing $d_p(x,y)$ covers most of the image area of the x,y fracture plane of Figure 2A. The spatial variation of $d_p(x,y)$ indicates that $\psi(x,y)$, the angle between the fracture plane normal and the normal, **z**, to bulk layers passing through the fracture plane, depends on position in the plane (see Figure S1). For a uniform periodic bulk layering of spacing d, the spacing $d_p$ in a plane intersecting having a normal making an angle $\psi$ with respect to the layer normal **z**, is $d_p = d/\sin\psi$, so that the smallest in-plane period is closest to the bulk spacing at $\psi = 90º$. The persistent theme of intersecting sets of nested rings of layers reveal a three dimensional (3D) structure of fluid, quasi-periodically spaced layers in the form of focal conic domains. Quite remarkable are the lines corresponding to the conic sections defining the domains (circles), identified by rows of cusps in the layers (Figure 2B) (*17*). In such cases, the conic sections are in the fracture plane yielding sets of rings that run down to very small radii (Figure 2C), and the layers pass through the fracture plane at $\psi \approx 90º$ (*17*). Therefore, in these domains the in-plane spacing $d_p(x,y)$ can be taken to be the bulk layer spacing, d, which we find by direct measurement on the images and Fourier analysis to be d = 8.3 ± 0.2 nm at T = 90ºC in the thermotropic NTB phase.

The FFTEM image of a 2D slice of 3D focal conics exemplified by Figure 2A exhibits unique FFTEM imaging characteristics in comparison to those of fluid layered liquid crystal systems such as chiral nematics, smectics, and columnar phases, as follows. In FFTEM of the 1D and 2D ordering of smectic and columnar LCs, respectively, fracture planes have a strong tendency to follow the interfaces between layers, so that the images are dominated by smooth layer interface surfaces rather than by layer edges (see Figures S2, S3). In columnar phases the translational ordering within the layers is then evident in the smooth layer surfaces (Figure S3). The FFTEM images of CB7CB show none of these features, but rather exhibit a strong tendency to fracture in planes nearly normal to the layers. We take this observation to be evidence for the translational symmetry of the layering, i.e., for reduced confinement of the fracture due to the absence of distinct layer interfaces on the molecular scale. Even in Figure 2B, where the layers reorient to be nearly parallel to the fracture plane, the layer steps remain rather indistinct compared to those of typical smectics (Figures S1,S2). The proposed twist-bend nematic is basically a network of overlapped dimers without weak interfaces parallel to the layers for a fracture plane to find. In contrast, a fracture plane normal to the layers requires only the breaking of side-to-side molecular contacts. Additionally, the fractures show no evidence for 2D ordering since, apart from the layer modulation, the fracture faces are quite irregular, even where the bulk layers are normal to the fracture plane. The distinction between the fracture of end-to-end contacts vs. side-to-side contacts enables freeze fracture to visualize director distortion in nematics (*18,19*), including the helical ordering of chiral nematic (*18,20*) and blue phases (*18,20,21*) on the >100 nm length scale. The present results extend the application of the FFTEM technique down to the 8nm scale in nematics, where it appears to be quite effective.

A selection of images from replicas of the NTB phase, quenched from T = 90, 95, 100, and 105 ºC (Figures S1, S4-S15) shows that at T = 90 and 95 ºC, virtually all areas of the replicas that are recovered and imaged exhibit the ~8nm stripes. At T = 100ºC patches without stripes begin to appear, coexisting with striped domains and exhibiting only the larger-scale roughness (Figures S9,S10). Figure 4 compares this stripeless state to examples of the stripe tex-ture at higher magnification. At T = 105 ºC no stripes are observed on the repicas: only the larger-scale roughness is found, showing that the stripe pattern is a property of the NTB phase. Remarkably, the NTB layering can also be observed near room temperature (T = 29 ºC) in the deeply supercooled NTB glassy state, i.e., when the sample is cooled over a 1 minute period from T = 95 ºC to T = 29 ºC and then quenched (Figures S11-S13). The stripe contrast and patterns exhibit only subtle changes over the entire range of T where they are found, including that the bulk layer spacing shows little change over the full temperature range of the thermotropic and supercooled NTB phase (Figure 3).

Pursuit of the idea that the stripe patterns in the FFTEM images are visualizing the TB helix in the NTB phase requires determining the relationship between the bulk stripe period determined from the FFTEM, d, and the pitch of the TB helix, $p_{TB}$, defined in Figure 1. Considering the predominant case in the FFTEM that the helix axis is parallel to the fracture plane or nearly so, Figure 1B shows that, while at each z within the $2\pi$ azimuthal reorientation period defining $p_{TB}$ the helix has a distinct azimuthal orientational state, the cases of $n(z)$ being either parallel or normal to the fracture plane actually occur twice within each pitch. If, for example, the each stripe corresponded to a level where the molecules were simply parallel to the surface, then the TB helix pitch would be $p_{TB} = 2d$, twice the FFTEM stripe period. However in such a case alternate stripes would correspond to alternate tilt of $n$ relative to $z$ in the fracture plane, say even stripes having the tilt to the left to someone looking down on the surface and odd layers having tilt to the right. Given the broad range of fracture circumstances evident in the FFTEM images shown in this report, such an alternation in tilt would inevitably under some conditions lead to an alternation in the appearance of the even and odd layers. The stripe pattern would then exhibit a cell doubling, indicating a unit cell of two stripes and a pitch of dimension 2d. However, we have never found any evidence for such a cell doubling in any of the hundreds of FFTEM images of the CB7CB NTB phase that we have studied, including the ones presented here. In the Fourier transforms of the stripe pattern intensity, such a cell doubling would show up as a subharmonic Bragg reflection at $q \approx 0.5*(2\pi/(8.3nm))$, at half the wavevector spacing of the fundamental stripe period. We have searched our images for both stripe alternation and the half-order reflection and have found neither, a result evident in the images and Fourier transforms presented here. From this we conclude that in every case the fracture process is uniquely relating azimuthal orientation in the TB helix to topographic height on the fracture face, so that the FFTEM stripe spacing is the TB helix pitch, i.e., $d = p_{TB} \approx 8.3$ nm and $q_{TB} \equiv 2\pi/p_{TB} \approx 0.76$ nm$^{-1}$.

The observation of this bulk $d \approx 8.3$ nm periodicity in the thermotropic NTB phase motivated and guided a synchrotron x-ray scattering search for a corresponding x-ray Bragg reflection. In these experiments samples were unoriented "powders" of CB7CB or reference material 8CB in 1mm diameter capillaries, with scattered intensity vs. wavevector I(q) measured with a scanning diffractometer (15), with results shown in Figures 5 and S13. I(q) for several sample conditions, including the starting room temperature crystal phase, and the N and NTB LC phases obtained upon cooling are compared in Figure 5 with a scan of the fundamental smectic A lamellar reflection of 8CB at T = 24 ºC with the same experimental conditions and counting time.

This search began by considering what scattering characteristics were to be expected from the $d \approx 8.3$ nm modulation found in FFTEM: (**i**) *wavevector* - The FFTEM data, which indicates that wavevector $q_{TB} = 2\pi/(8.3\text{nm}) \approx 0.76$ nm$^{-1}$ and its harmonics, indicated by the vertical yellow bars in Figure 5, would be of most interest. (**ii**) *Bragg peak intensity* – LCs exhibit non-resonant x-ray diffraction because of variation in electron density. The 8CB SmA Bragg peak, included in Figure 5, can serve as an intensity standard because computer simulation of SmA 8CB (*22*) shows that at T = 24 ºC the peak-to-peak modulation of mass density in the the smectic A wave is from $\rho = 0.87$ g/cm$^3$ to $\rho = 1.13$ g/cm$^3$, so that the 8CB smectic reflection is from a 1D wave of electron density with a fractional modulation amplitude of $\delta\rho/\rho \approx 0.25$. With this information, comparison of the intensity of Bragg scattering from the NTB phase with that of SmA 8CB can then be used to determine a maximum possible fractional electron density in the NTB. (**iii**) *Bragg peak lineshape* – The Bragg peak shown in Figure 5 for SmA 8CB has nearly the shape of the diffractometer resolution function (full width at half height (FWHH) $\delta q = 0.005$ nm$^{-1}$)), enhanced slightly in the tails by the thermal undulations of the smectic layers via the Landau-Peierls (LP) effect, which limits smectic order in 3D to be quasi-long-ranged (QLRO) (23). The NTB modulation should also exhibit only QLRO, but the well-defined nature of the focal conics found in DTLM and in the NTB phase FFTEM images, and the weak layer positional decorrelation in some FFTEM images, for example Figure S13, suggest that, as in smectics, the Landau-Peierls disordering should be weak. Direct evidence for this is available from data on the $SmC_\alpha$ phase, which exhibits an precession of period ~10nm in the azimuthal orientation of molecules on the SmC tilt cone to form a helix of just the TB type and pitch under consideration for the NTB (14,24,25). Resonant x-ray Bragg scattering from the $SmC_\alpha$ TB director helix shows FWHH values of $\delta q \sim 0.02 nm^{-1}$ (25), a value we could also expect for the NTB phase since the TB helices in the two cases are of comparable elasticity, and our diffractometer has sufficient resolution. We therefore would expect a sharp peak from the NTB modulation if it scatters. (**iv**) <u>background</u> – Background comes from stray scattering from the beam path. The high angular resolution of the diffractometer rejects nearly all of the intensity of diffuse features in the sample scattering, but passes essentially all of peaks that are near the resolution-limit in width. This is an ideal situation to search for weak scattering from a well-ordered modulation like that found in the FFTEM.

In fact, the x-ray scans show no evidence for Bragg reflection near $q_{TB}$ or any of its harmonics in the NTB phase. An upper limit on possible NTB Bragg intensity can be established as follows. The background at the first harmonic $I_b(q) = 48.0 \pm 0.2$ counts can be determined precisely by fitting a low order polynomial to the background over a broad range of q, as shown in Figure 5. Subtracting the fitted function from the data leaves only its shot noise, ~ ±7 counts at $q \sim 0.76$ nm$^{-1}$ as the uncertainty on each point (Figure 5B). Figure 5A shows these data smoothed over 10 adjacent points, such that any peak would have the width of that observed for the $SmC_\alpha$ helix, as discussed above ($\delta q = 0.03$ nm$^{-1}$). This smoothing leaves fluctuations of ~±2 counts for each point (Figure 5A), and ~ ±2*10 = ±20 counts in the area of local fluctuations. The area of the 8CB SmA peak is 60,000 counts over background, so that these shot noise fluctuations are equivalent to a ratio of TB reflection peak area to that of SmA 8CB of 20/60,000 which, since $\delta\rho/\rho \propto \sqrt{I(q)}$, corresponds in turn to fractional TB phase electron density modulation fluctuations of $(\delta\rho/\rho)_{fluc} \sim 0.25/\sqrt{3000} = \sim\pm0.005$. The absence of any features in the scattering above this noise level therefore limits the amplitude of the density modulation wave of period p to $\delta\rho/\rho \sim <0.005$ in the NTB phase.

Motivated by these observations and by recent examples of the very successful description of the physical properties of the nematic and smectic A phases of cyanobiphenyls by atomistic computer simulation (22,26,27), we have carried out molecular dynamic (MD) simulations of the ground state structure of the nematic phases of CB7CB, and, for comparison, of the nematic phase of CB6CB (Figure 6A,B). These simulations, which combine a widely-tested fully-atomistic force field (28), with advanced MD techniques (29, 30,15), were set up in an orthorhombic 5.6nm* 5.6nm * 8nm cell with periodic boundary conditions and the cell dimension along **n** set to $L_z \sim 8$ nm, as suggested by the FFTEM data. Initially, simulations were conducted with a biasing potential that aligned the mesogens along **z**, achieved by applying weak forces to the cyano groups at the end of each molecule, pulling them in the opposite (+z and -z) directions. As a result, the initially equilibrated configurations were in a well-defined nematic phase (N) with the nematic director **n** aligned along **z**. Subsequently, the biasing potential was turned off and each system was simulated in the $NP_zT$ ensemble in the 370 - 410K temperature range with the box dimensions allowed to fluctuate to achieve atmospheric pressure on all faces.

The comparison of nematic ordering formed at 370K clearly highlights the difference between the odd and even number of carbons in the alkyl spacer. For the CB6CB, simulations over 15ns equilibrated a typical N structure as illustrated in Figure 6A and C. However, the CB7CB system, which initially had a similar nematic structure due to the above-mentioned biasing during equilibration, within a few nanoseconds after the biasing was removed showed a clear spontaneous deformation into a helical structure having its axis oriented along the original nematic director, **z**, as illustrated in Figures 6B and D. The highlighted mesogens exhibit a complete period of conical helical twist on a $p_{TB} \sim 8$nm length scale. Equilibration will adjust the box length $L_z$ to match the repeat distance of a periodic structure along z if the latter is initially sufficiently close to $L_z$. Extended equilibration with $L_z$ initially in the range 8.3nm < $L_z$ < 7.9nm relaxed to $L_z$ = 8.1nm, determining $p_{TB}$ = 8.1nm as the TB period for this simulated molecular system.

Our analysis shows that few of the molecules in the simulated N and NTB phases have the lowest energy all-trans conformations shown in Fig. 6A and B. Nevertheless, calculation of the angle (β) between relative orientation of two CB units on the same molecule (Figure 6) shows that in the NTB phase the CB7CB molecules maintain a significant average bend with $\langle β \rangle$ = 133°, while in the N phase the CB6CB molecules are almost linear, with $\langle β \rangle$ = 166°. The average tilt angle of the CB units, i.e., of the half molecular director, **n**, relative to **z**, was found to be $\langle θ \rangle$ = 19° for the N state of CB6CB, and $\langle θ \rangle$ = 33° for the NTB state of CB7CB.

The striking lack of temperature dependence of TB helix pitch $p_{TB}$ in the NTB phase, shown in Figure 3, and the first order nature of the N-NTB transition found in a variety of experiments, suggests that the TB helix is a principally enthalpically stabilized structure and calls into question the picture of the N-NTB transition as being driven by a negative bend Frank constant. Rather these features suggest specific molecular pairing motifs, perhaps into living-polymerized-like helical chains that can lock in a particular pitch, as suggested by Figure 2b of Dozov's paper (*12*) and Figure 1D above. As a result, variation of chemical structure, for example MD study of the dependence of $\langle θ \rangle$ and $p_{TB}$ on n in the n-odd CBnCB homologous series, should be particularly interesting. A Landau model providing a first order transition to a state of preferred bend, driven by a Frank energy having a term linear in the bend deformation, of the type suggested by Meyer (*11*), may be a better picture of the transition to the TB state in CB7CB.

The external force-induced stabilization of the N phase in CB7CB reported above is a quadrupolar analog of the "conical helix to field-polarized" transition in the helimagnet MnSi (*31*). Our simulations make it clear that inducing the N phase in simulation by any of a variety of possible couplings tending to align the CB end groups along **z** should be a powerful way of exploring nature of the N-NTB phase transition. This relationship with helimagnets further sug-gests the possibility of a TB nematic analog of the MnSi "A" phase wherein, at higher temperature the conical helix reorganizes into a 2D hexagonal blue phase lattice of double twist cylinders (*31*).

Another aspect of the NTB structure that can be discussed in the light of these observations and the implied local chirality of the TB helix, is macroscopic chirality. Currently there is no direct information on the dimensions of homochoral domains in the CB7CB NTB phase in absence of external perturbation, although they must be large enough for the local handedness to persist in the NMR timescale (~1 msec) (*4*), and electric field treatment generates macroscopic conglomerate domains (*9*). The lower magnification FFTEM images presented here cover areas of ~ 2μm x 2μm square, and in such images we have not found any clear situations where the handedness is clearly changing sign. Qualitatively, we would expect such a conglomerate domain boundary to be marked by a disappearance of the layers if it were running normal to the layers, or by an irregularity in the layer spacing if running parallel to the layers. Absence of such features indicates that the as-grown conglomerate domains are of multi-micron dimension, approaching being macroscopic with respect to optical probes and suggesting that they may exhibit optical rotation (OR) or circular dichroism. Having such large domains in the field-free state may account for the observation of electric field-induction of truly macroscopic conglomerate domains, of dimensions > 100 μm (*9*).

Nevertheless, the conglomerate domains of the helix in the as-grown NTB phase exhibit little detectable optical rotation (OR) in absence of applied electric field, and barely detectable OR in the large conglomerate domains obtained by field treatment, although the latter show an electroclinic-like chiral field-induced reorientation of the uniaxial optic axis (*32*). The chiral optical effects to be expected in macroscopic NTB domains, such as optical rotation (OR), can be estimated on the basis of the study of short pitch SmC helices (*33*) the B2 phases of bent core molecules (*34,35,36*). The NTB helix of 8 nm pitch is nearly identical in structure to the effective director helix in the $SmC_AP_A$ phase (pitch = 70nm) (*34,35*), and should exhibit comparable optical properties. Optical characterization of focal conic domains of a typical $SmC_AP_A$ yields an OR = 0.05°/μm (36), in agreement with optical modeling of TB structures (35), and shows that this OR is extremely difficult to detect in the presence of the birefringence of the TB helix. Modeling also shows that the OR in the homeotropic orientation of the NTB TB helix, having the helix axis normal to the surfaces cell plates, should be even smaller, by several orders of magnitude (*33,34*).

A conical NTB helix of the sort sketched in Figure 1C is accompanied by a commensurate helical precession of a polarization density field, locally always normal to both **n** and **z**, also shown in Figure 1C. The field-induced chiral electroclinic optic axis reorientation can be understood as a coupling of applied field to this helielectric polarization field **P(r)**, a resulting distortion of the helix by local rotation of **P(r)** toward the field direction, and a consequent tilt of the average director orientation, $\omega \propto \mathbf{z} \times \mathbf{E}$ (*37, 38*). This effect is weak in the NTB phase because $p_{TB}$ is so small (38), but, by the same argument, the converse effect, the induction of macroscopic polarization by flow (*39*), ought to be large.

*Acknowledgements:*

The authors thank Leo Radzihovsky for stimulating discussions. This work was supported by NSF MRSEC Grant DMR 0820579.

*Author Contributions*:

D.C. carried out the FFTEM experiments. J.P., E.K., and D.W. synthesized CB7CB. J.H., D.B., and M.G. carried out computer simulations. N.C., A.K., and J.E.M. carried out optical microscopy. Y.S. performed x-ray diffraction experiments. D.C., D.B. and N.C. analyzed data. N.C., D.C., and J.E.M. wrote the paper. N.C. and D.W. directed the project.

*Figure 1:*

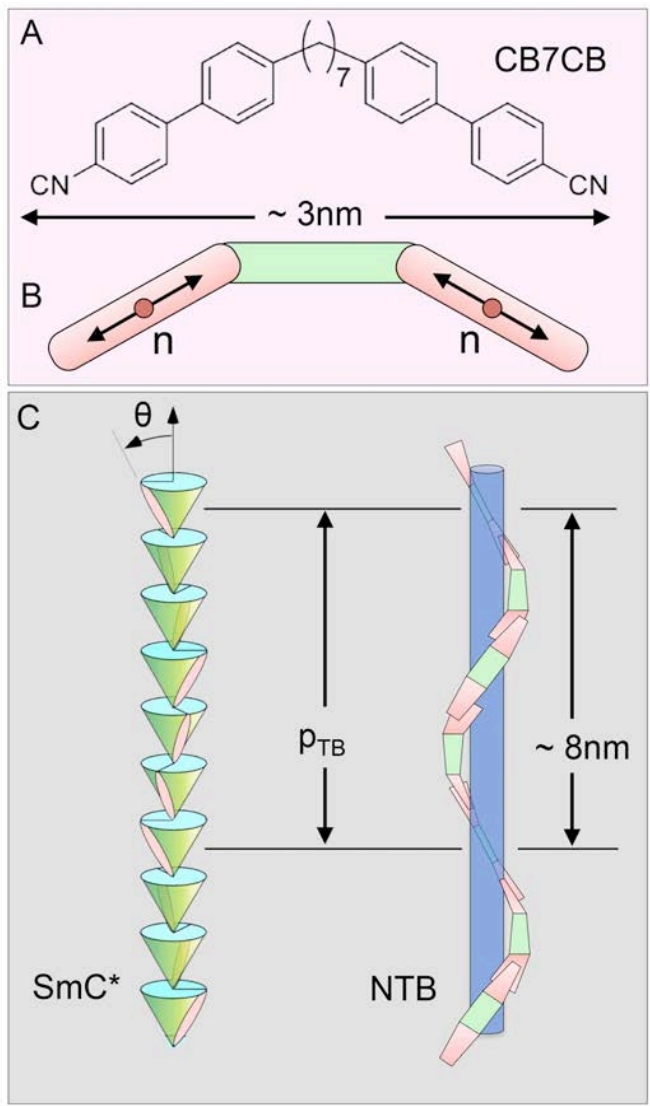

*Figure 1* – (A) Structure of CB7CB [4',4'-(heptane-1,7-diyl)bis(([1',1"-biphenyl]-4"-carbonitrile))]. The end-to-end length of its van der Waals surface is ≈3.0 nm in the all-trans molecular configuration. (B) Schematic structure of CB7CB, which can be viewed as having three parts, each ~1nm in length: two rigid end groups connected by a flexible spacer. The nematic director field, **n(r)**, is the local average orientation of the long axes of the rigid end groups, located by the red circles. (C) Schematic illustration of (TB) helices in the layered chiral smectic C (SmC*) phase and in the layerless nematic twist-bend (NTB) phase. The NTB drawing is a qualitatively correct representation of the TB structure found in CB7CB, in which any interval along the helix of single pitch length, $p_{TB}$ ~ 8nm, is made up on average of 4 overlaps of 1nm long rigid ends and 4 intervals of 1nm long flexible spacers.

*Figure 2:*

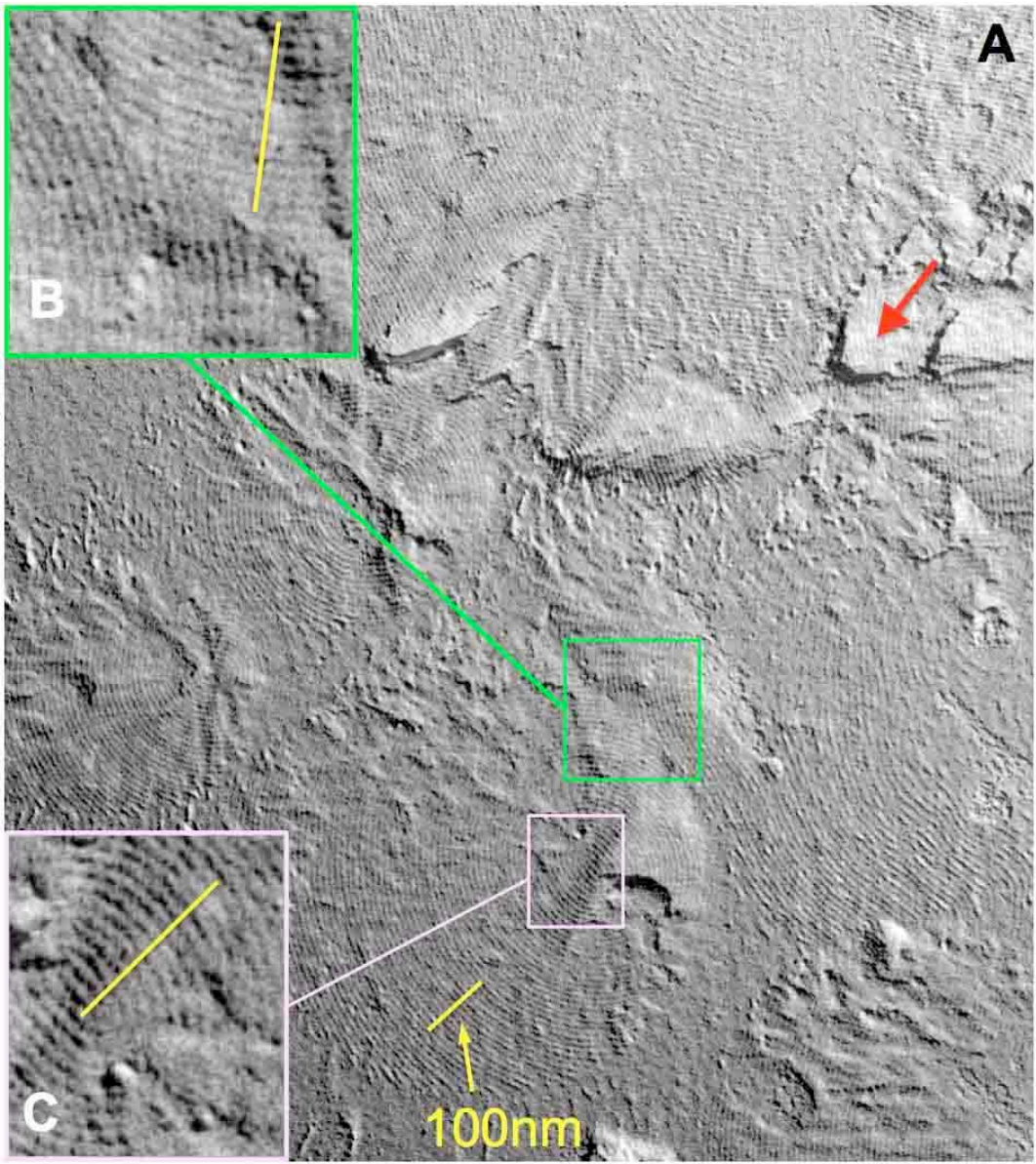

*Figure 2* – Freeze fracture TEM image of CB7CB quenched from the NTB phase T = 95 ºC, with the Pt shadowing direction indicated by the red arrow. The fracture face exhibits a 2D texture of curved periodically arrayed layers indicative of a bulk fluid layered structure of 3D focal conics, with conic section lines parallel to (B) and normal to (C) the fracture surface. The bulk layer spacing is that observed in (C), where the layers are normal to the fracture plane (Figure S1). The evident tendency for the fracture plane to run normal to the layers distinguishes TB layering from smectics, where fractures are mostly between smectic layering (Figures S2, S3).

*Figure 3:*

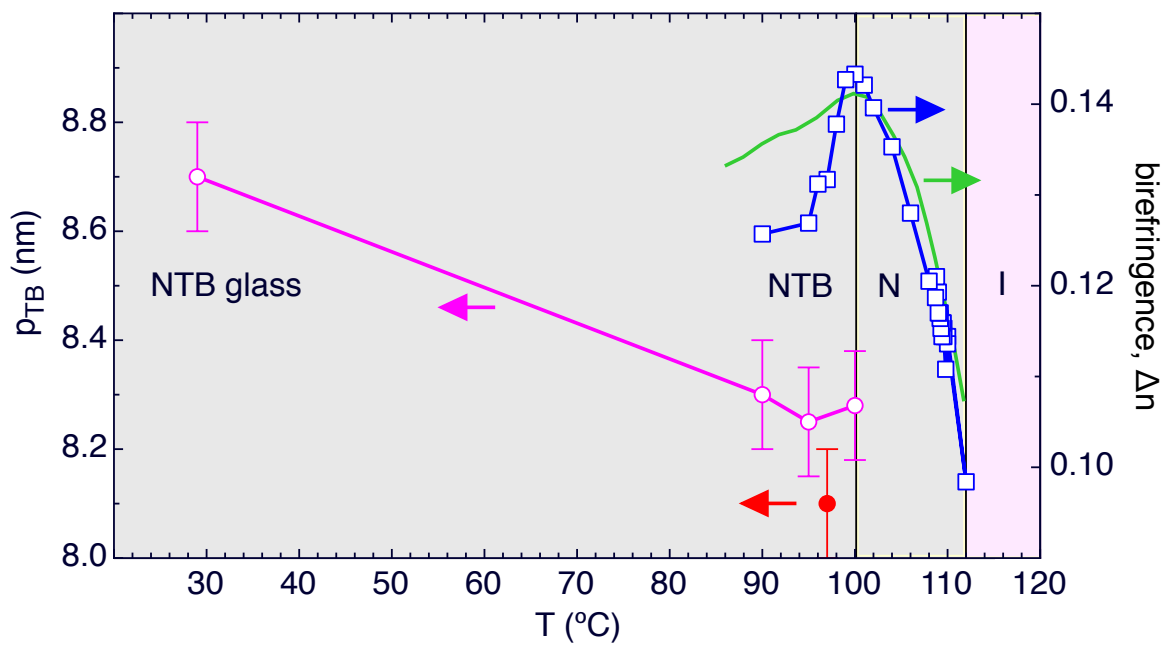

*Figure 3* – Plot of the temperature dependence the helix pitch of the NTB phase, $p_{TB}(T)$, and of the birefringence, $\Delta n(T)$ (□), of CB7CB in the N and NTB phases. The pitch from the FFTEM images (○) exhibits little variation with T. The solid circle (●) gives $p_{TB}$ from the MD computer simulation of Figure 6. The birefringence grows with decreasing T in the N phase, as is typical. The solid green line is proportional to the DMR splitting-determined orientational order parameter of deuterated 8CB solute in CB7CB (*8*), and is proportional to $\Delta n(T)$ in the N phase. In the NTB phase $\Delta n(T)$ decreases with the onset of biaxial ordering, in a fashion different from the DMR spltting.

*Figure 4:*

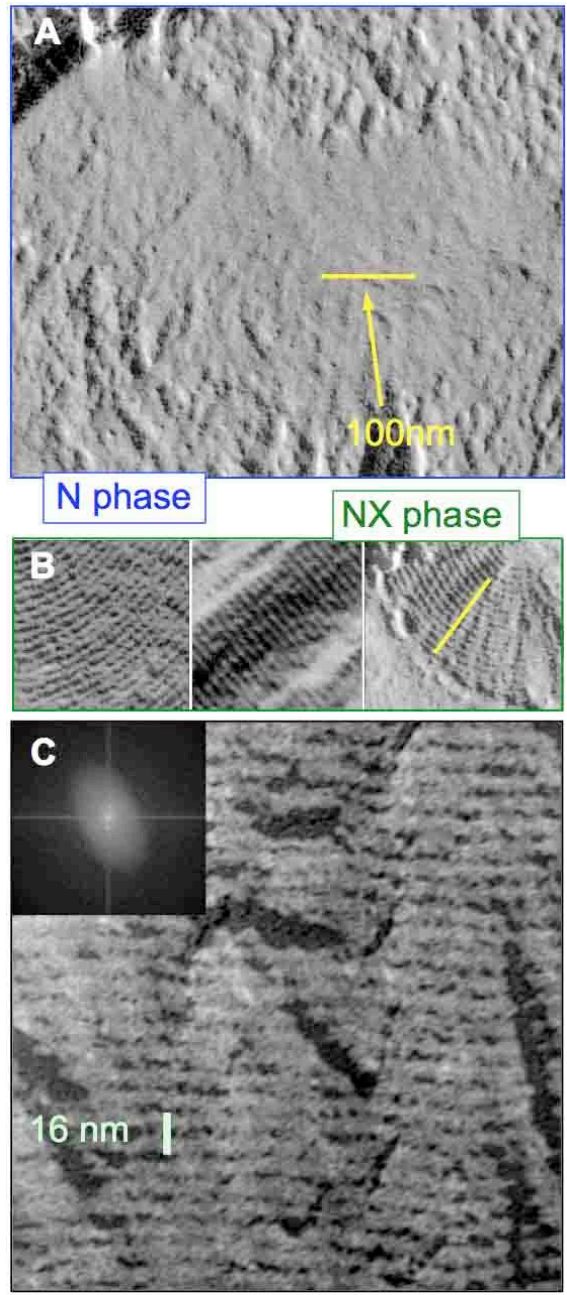

*Figure 4* – (A,B) Comparison of FFTEM images of the N phase quenched from T = 105 ºC and of the NTB phase quenched from T = 95 ºC, showing that the fracture surface periodicity disappears in the N phase. (C) Higher magnification FFTEM of the CB7CB NTB phase, showing the inherent ~ 0.5nm FFTEM resolution. This image exhibits, in addition to the TB layering, a chiral aperiodic intra-layer modulation of tilted nm-scale features that produces the tilted halo in the image Fourier transform (inset). This chiral structure may result from the chirality of the TB helix.

*Figure 5:*

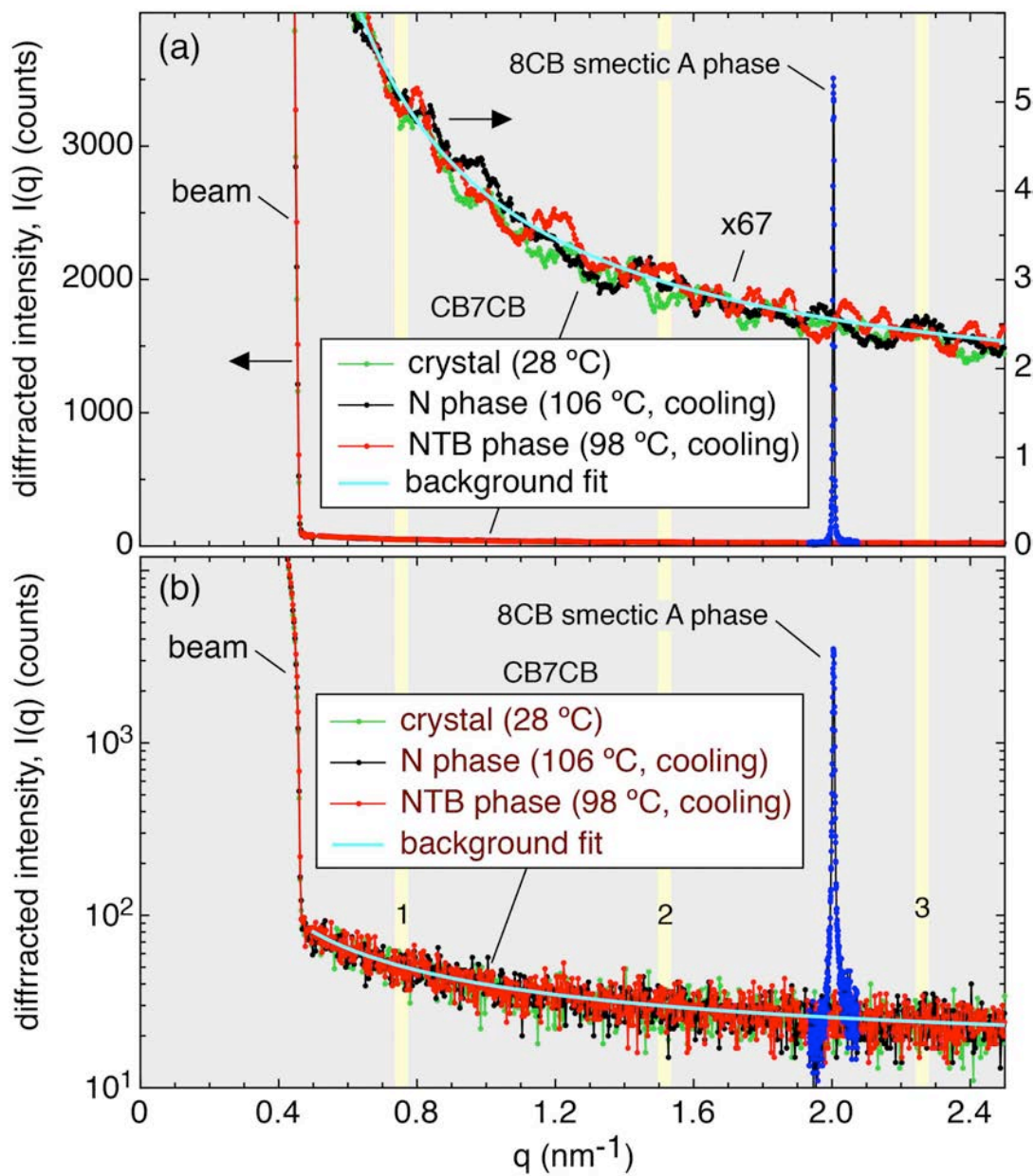

*Figure 5* – If there is electron density modulation (EDM) associated with the periodic layering found in FFTEM of CB7CB, it should generate sharp x-ray diffraction (XRD) peaks at wavevector $q \sim 2\pi/8$ nm$^{-1}$ and its harmonics (yellow bands). Here, in a search for this scattering, synchrotron-based powder XRD from CB7CB is compared with that from the smectic A layering in 8CB, which computer simulation shows to have a fractional EDM of 0.25. Subtraction of background (cyan line) leaves only the shot noise from the background to limit detectability of a peak. No peaks above this limit are found, indicating that the fractional EDM in any TB scattering structure in BC7CB must be less that 0.005.

*Figure 6:*

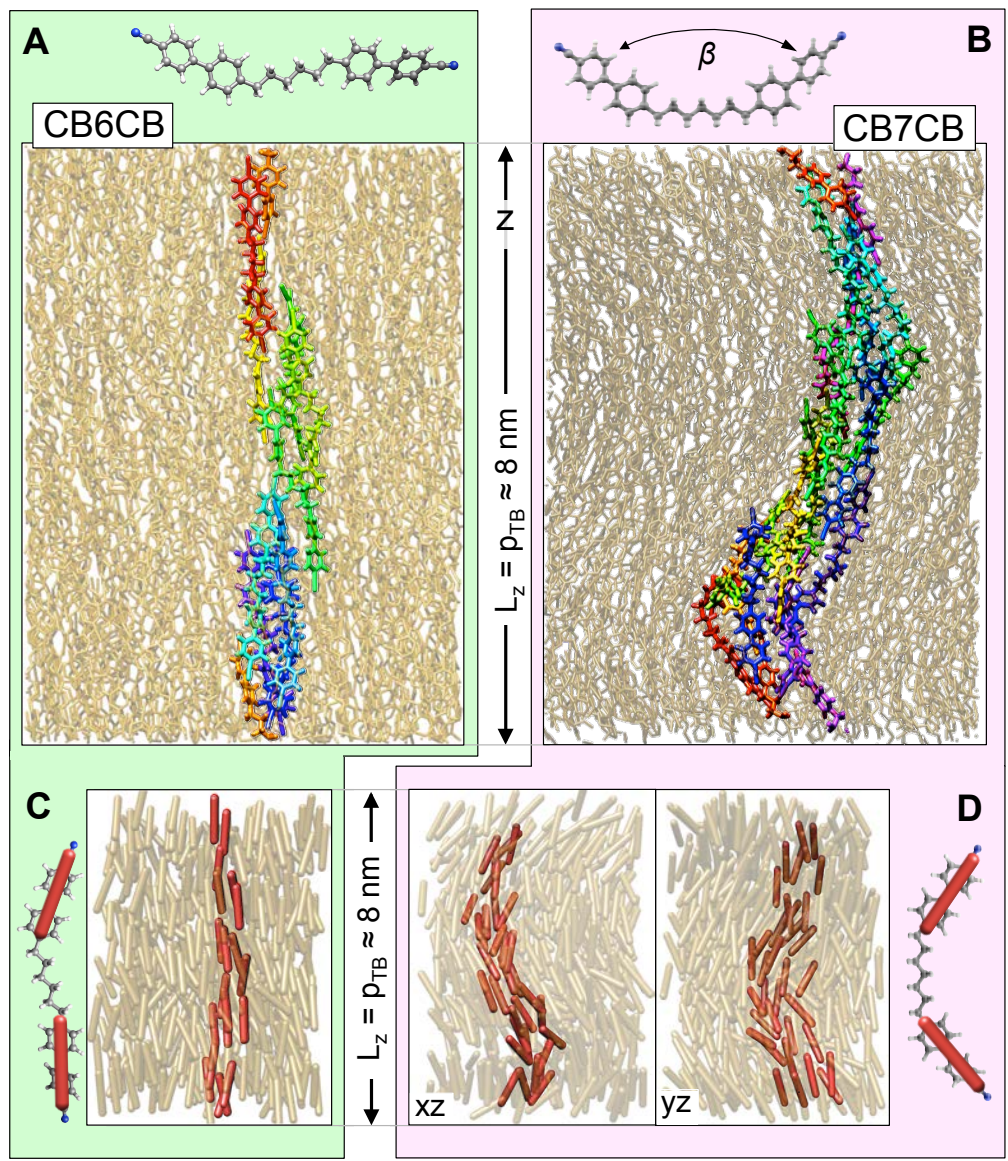

*Figure 6* – Fully atomistic molecular dynamic simulations of equilibrium nematic phases at T = 370K in CB6CB, a linear molecule (A), and CB7CB, a bent molecule (B), showing a view of a nominally 5.6 * 5.6 * 8.0 nm periodic box. Initial equilibration is carried out with opposed forces in the z-direction on the molecular ends enforcing a form of field-induced nematic monodomain. As these forces are relaxed CB6CB remains an nematic with **n** along **z** (A), whereas CB7CB relaxes into a conical TB helix, of pitch $p_{TB}$ = 8.1nm and cone angle θ = 33º (B). The periodic box adjusts in length along z to match $p_{TB}$. (C,D) Example configurations showing explicitly the orientations of the biphenyl molecular groups that determine the director field and their clear twist-bend ordering in CB7CB.

*Supplementary Material:*

<u>*A Twist-Bend Chiral Helix of 8 nm Pitch*</u>
<u>*in a Nematic Liquid Crystal Phase of Achiral Bent Molecular Dimers*</u>

Dong Chen[1], Jan H. Porada[2], Justin B. Hooper[3], Arthur Klittnick[1], Yongqiang Shen[1],
Eva Korblova[2], Dmitry Bedrov[3], David M. Walba[2],
Matthew A. Glaser[1], Joseph E. Maclennan[1] and Noel A. Clark[1]

[1] Department of Physics and Liquid Crystal Materials Research Center,
University of Colorado, Boulder, CO 80309-0390

[2] Department of Chemistry and Biochemistry and Liquid Crystal Materials Research Center,
University of Colorado, Boulder, CO 80309-0215

[3] Department of Materials Science and Engineering, University of Utah, Salt Lake City, UT 84112, and Liquid Crystal Materials Research Center, University of Colorado, Boulder, CO 80309-0215

SUPPLEMENTARY NOTES

*1. Synthesis of CB7CB*

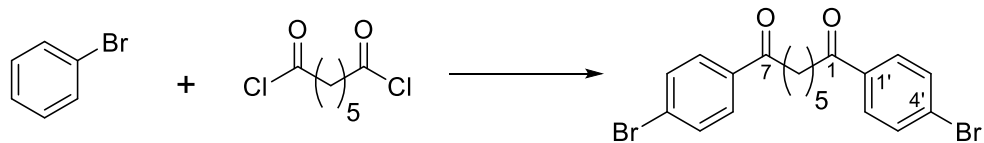

1,7-bis(4'-bromophenyl)heptane-1,7-dione: A solution of 5.00g (25 mmol) pimeloyl chloride in 5 ml (45 mmol) bromobenzene was slowly dropped into an ice cooled and stirred suspension of 7.07g (53 mmol) aluminum trichloride in 30 ml (285 mmol) bromobenzene under argon. The reaction mixture stirred for 12 hours and allowed to warm to room temperature, before it was poured on ice water (200 ml). The organic phase was extracted with dichloromethane and washed once with sodium hydrogen carbonate solution, once with water and was dried over magnesium sulfate. The solvent and excess bromobenzene was removed in vacuum and the product was recrystallized from ethanol yielding 10.07g (23 mmol, 92%) of colorless crystals.

$^1$H NMR (300 MHz, CDCl$_3$) δ 7.81 (d, J = 8.8 Hz, 4H; 2'-H and 6'-H), 7.60 (d, J = 8.8 Hz, 4H; 3'-H and 5'-H), 2.95 (t, J = 7.3 Hz, 4H; 2-CH$_2$ and 6-CH$_2$), 1.87 – 1.69 (m, 4H; 3-CH$_2$ and 5-CH$_2$), 1.56 – 1.39 (m, 2H; 4-CH$_2$).

$^{13}$C NMR (75 MHz, CDCl$_3$) δ 199.26 (s; C-1 and C-7), 135.82 (s; C-1'), 132.04 (d; C-3' and C-5'), 129.71 (d; C-2' and C-6'), 128.26 (s; C-4'), 38.37 (t; C-2 and C-6), 28.97 (t; C-4), 24.02 (t; C-3 and C-5).

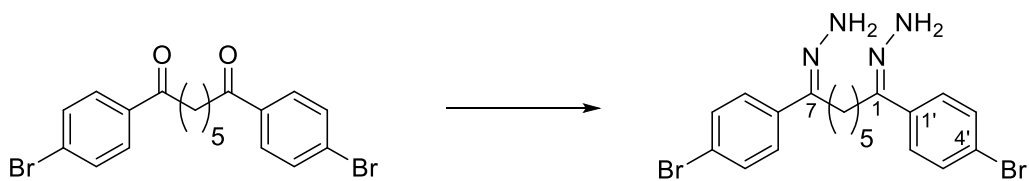

1,7-bis(4'-bromophenyl)-1,7-dihydrazonoheptane: 10.00g (22.8 mmol) 1,7-bis(4-bromophenyl)heptane-1,7-dione were dissolved in 70 ml abs. ethanol and 5.5 ml hydrazine hydrate were added. The solution was refluxed for 12 hours, cooled to room temperature and was further cooled in the freezer. The precipitate was filtered off, giving 6.244g (13.4 mmol, 59%) of 1,7-bis(4-bromophenyl)-1,7-dihydrazonoheptane as slightly yellow powder.

$^1$H NMR (300 MHz, CDCl$_3$) δ 7.47 (m, 8H; 2'-H, 3'-H, 5'-H and 6'-H), 5.42 (s, 4H; NH$_2$), 2.70 – 2.38 (m, 4H; 2-CH$_2$ and 6-CH$_2$), 1.73 – 1.31 (m, 6H; 3-CH$_2$, 4-CH$_2$ and 5-CH$_2$).

$^{13}$C NMR (75 MHz, CDCl$_3$) δ 149.19 (s; C-1 and C-7), 137.51 (s; C-1'), 131.58 (d; C-3' and C- 5'), 127.21 (d; C-2' and C- 6'), 122.27 (s; C-4'), 30.17 (t; C-4), 25.12, 24.92 (t; C-2, C-3, C-5 and C-6).

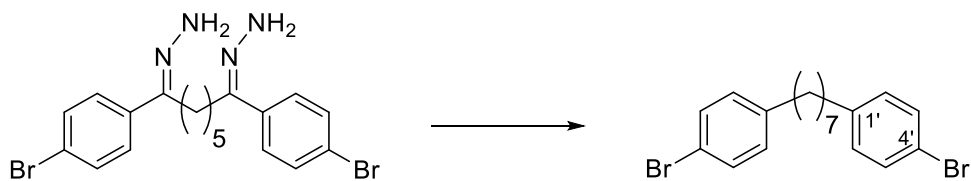

1,7-bis(4'-bromophenyl)heptane: 6.06g (13.0 mmol) of 1,7-bis(4-bromophenyl)-1,7-dihydrazonoheptane were solved in 100 ml dry toluene and 7.29g (65.0 mmol) of potassium tert.-butoxide were added. The solution was refluxed under argon for 48 hours, cooled to room temperature and poured into 100 ml ice cooled 1M HCl. The organic layer was separated and the aqueous phase was extracted with dichloromethane. The combined organic phases were washed once with water, once with brine and were dried over magnesium sulfate. The solvent was removed and the crude product was recrystallized from ethanol, giving 3.95g (9.6 mmol, 74%) of 1,7-bis(4-bromophenyl)heptane as colorless crystals.

$^1$H NMR (300 MHz, CDCl$_3$) δ 7.39 (d, J = 8.3 Hz, 4H; 3'-H and 5'-H), 7.04 (d, J = 8.4 Hz, 4H; 2'-H and 6'-H), 2.55 (t, J = 7.7 Hz, 4H; 1-CH$_2$ and 7-CH$_2$), 1.65 – 1.50 (m, 4H; 2-CH$_2$ and 6-CH$_2$), 1.42 – 1.22 (m, 6H; 3-CH$_2$, 4-CH$_2$ and 5-CH$_2$).

$^{13}$C NMR (75 MHz, CDCl$_3$) δ 141.84 (s; C-1 and C-7), 131.38 (d; C-2' and C- 6'), 130.29 (d; C-3' and C- 5'), 119.40 (s; C-4'), 35.44 (t; C-1 and C-7), 31.38 (t; C-2 and C-6), 29.38 (t; C-4), 29.16 (t; C-3 and C-5).

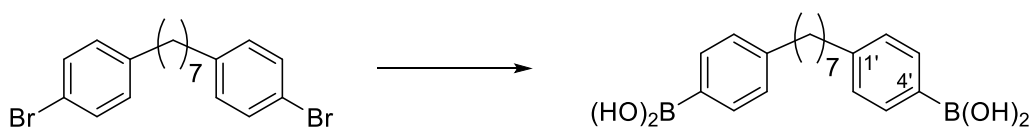

(heptane-1,7-diylbis(4',1'-phenylene))diboronic acid: 3.77g (9.2 mmol) of 1,7-bis(4-bromophenyl)heptane were solved in 50 ml dry THF and cooled to -78°C. 12.5 ml (20 mmol) of n-butyl lithium (1.6M in hexane) were added slowly and the reaction mixture was stirred for 0.5 hours at -78°C. 4.5 ml (40 mmol) trimethyl borate were added and the mixture was allowed to warm to room temperature before 40 ml of 5M HCl were added. The solution was extracted with diethyl ether and the organic phase was washed twice with water, once with brine and was dried over magnesium sulfate. The solvent was evaporated and gave 3.07g (9.0 mmol, 98%) of crude (heptane-1,7-diylbis(4,1-phenylene))diboronic acid as colorless paste, which was used without further purification.

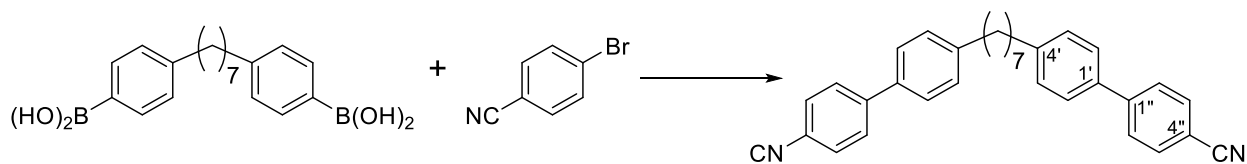

4',4'-(heptane-1,7-diyl)bis(([1',1''-biphenyl]-4''-carbonitrile)): 1.02g (3.0 mmol) (heptane-1,7-diylbis(4,1-phenylene))diboronic acid, 1.20g (6.6 mmol) 4-bromobenzonitrile and 2.07g (15.0 mmol) potassium carbonate were suspended in a mixture of 50 ml ethanol and 10 ml water and degassed by applying vacuum and flushing with argon under ultrasonic irradiation. Then 0.346g (0.3 mmol) tetrakis(triphenylphoshine)palladium(0) were given to the reaction mixture and it was refluxed for 12 hours. The solvent was evaporated in vacuum and 30 ml dichloromethane and 30 ml water were given to the residue. The organic phase was separated, washed twice with water, once with brine and was dried over magnesium sulfate. The solvent was evaporated and the crude product was purified chromatographically on silica with dichloromethane as eluent yielding 0.747g (1.64 mmol, 55%) of 4',4'-(heptane-1,7-diyl)bis(([1',1''-biphenyl]-4''-carbonitrile)) as a colorless solid.

$^1$H NMR (300 MHz, CDCl$_3$) δ 7.75 – 7.62 (m, 8H; 2''-H, 3''-H, 5''-H and 6''-H), 7.51 (d, J = 8.2 Hz, 4H; 2'-H and 6'-H), 7.29 (d, J = 8.5 Hz, 4H; 3'-H and 5'-H), 2.66 (t, J = 7.7 Hz, 4H; 1-CH$_2$ and 7-CH$_2$), 1.75 – 1.56 (m, 4H; 2-CH$_2$ and 6-CH$_2$), 1.46 – 1.32 (m, 6H; 3-CH$_2$, 4-CH$_2$ and 5-CH$_2$).

$^{13}$C NMR (75 MHz, CDCl$_3$) δ 145.68 (s; C-1''), 143.80 (s; C-4'), 136.58 (s; C-1'), 132.68 (d; C-3'' and C-5''), 129.29 (d; C-3' and C-5'), 127.57 (d; C-2'' and C-6''), 127.19 (d; C-2' and C-6'), 119.15 (s; CN), 110.66 (s; C-4''), 35.72 (t; C-1 and C-7), 31.48 (t; C-2 and C-6), 29.47 (t; C-4), 29.33 (t; C-3 and C-5).

## 2 – X-Ray Diffraction

X-Ray diffraction was carried out on beamline X10A of the National Synchrotron Light Source (NSLS), Brookhaven National Laboratory. This beamline has a Si 111 double monochromater, typically tuned to around 10 KeV. The sample is mounted in an Instec hot stage on a Huber 4-circle goniometer. The point detection arm is equipped with a Ge monochromater to select the direction of diffracted x-rays. The angular resolution, measured by scanning the detector arm through the attenuated direct beam is typically δq ~ 0.005 nm$^{-1}$, full width at half maximum (FWHM).

## 3 – Atomistic Computer Simulation

A non-polarizable version of the APPLE&P force field [28] was used in all simulations treating all atoms explicitly. Simulations contained 384 mesogens set up in three-dimensional, periodic orthorhombic simulation cells. All simulations were conducted using the integration scheme proposed by Martyna et al. [29] with frequencies of 10$^{-2}$ and 0.5 × 10$^{-3}$ fs for the thermostat and barostat control, respectively. Initially, the mesogens were set up on a regular low-density lattice and then the system was shrunk to the desired dimension, i.e. x and y dimensions about 5.6nm and 8nm for the z-direction. During this compression and the subsequent equilibration the simulations were executed with biasing potential that aligned the mesogens along the z-axis. The latter was achieved by applying weak forces to cyano groups at the end of each molecule pulling them in the opposite (+z and -z) directions. As a result, the initially equilibrated configurations were in a well-defined nematic phase (N) with the nematic director aligned along z-direction. Subsequently, the biasing potential was turned off and each system was simulated in the $NP_zT$ ensemble with the z-dimension allowed to fluctuate to achieve the atmospheric pressure in the system, while x and y dimensions were kept fixed. During these simulations the cell dimension in the z-direction fluctuated within ~1.0 Å. The final stress tensor was very close to zero indicating that, as expected for a liquid, no residual stresses remained in the system due to applied constrains on cell dimensions. Simulations were carried out in the 370 - 410 K temperature range. Production runs were over 20 ns during which no drift in the system energy or order parameters have been observed indicating a stationary behavior of the system at applied conditions.

Simulations were conducted with bond lengths constrained using the Shake algorithm [30] to utilize a larger time step. The long-range electrostatic forces were treated by the Ewald summation method. A multiple time step reversible reference system propagator algorithm [29] was employed with a time step of 0.5 fs for bonding, bending, and torsional forces, a 1.5 fs time step for nonbonded interactions within a 7.0 Å cutoff radius, and a 3.0 fs time step for nonbonded interactions between 7.0 and 11.0 Å and the reciprocal part of the Ewald summation.

The lowest energy conformation for an alkyl chain is the all-*trans* configuration for the backbone dihedrals which leads to noticeably distinct orientations of the CB half-molecular units, as indicated in Figure 6B and D. In the CB7CB the angle, β, between the two CB unit axes is substantially less than β = 180°, forming a banana-like conformation, while the CB6CB has an almost linear configuration. However, the energy difference between *trans* (180°) and *gauche* (±60°) dihedral orientations is only 0.5kcal/mol, which at 370K is less than $k_BT$ of energy and therefore plenty of gauche conformations are expected in a quite flexible alkyl spacer due to thermal fluctuations. Our analysis of dihedral populations shows that in the formed nematic phases the spacers of both mesogens have on average about 30% of torsions in the *gauche* state and therefore the probability to find all-*trans* configurations of alkyl spacers is small.

## Supplementary Figures

*Figure S1*

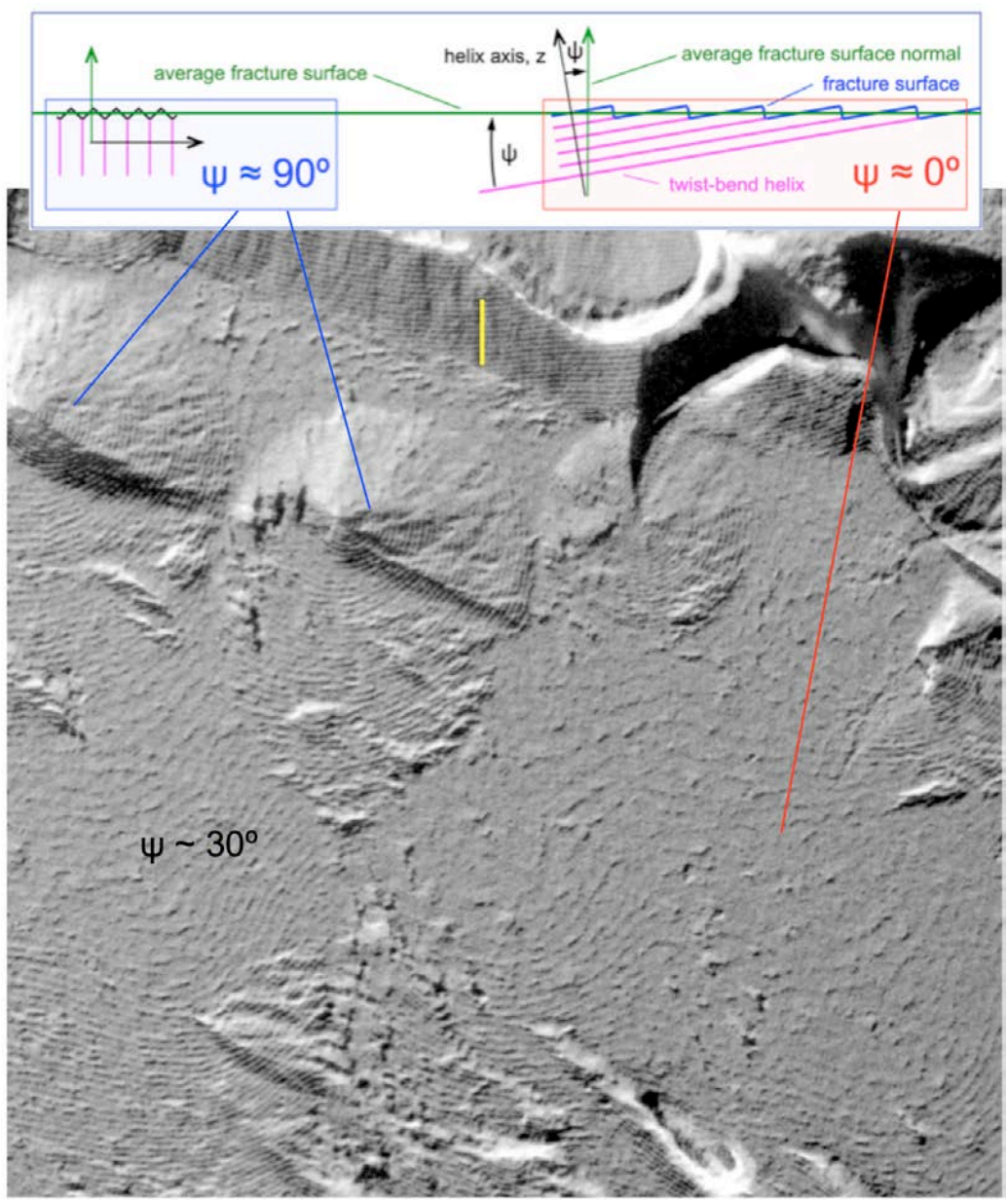

*Figure S1* - Fracture area of CB7CB at T = 90°C where the layer planes are nearly parallel to the image plane over the right central area of the image ($\psi \sim 0°$) and normal to the fracture plane in the focal conics ($\psi \sim 90°$). Where $\psi$ is small the fracture face becomes rather irregular, in contrast with the case of freeze fracture of smectic layered phases where the layer surfaces are smooth and exhibit with distinct layer steps (Figure S2). Scale bar (yellow) = 100nm.

*Figure S2*

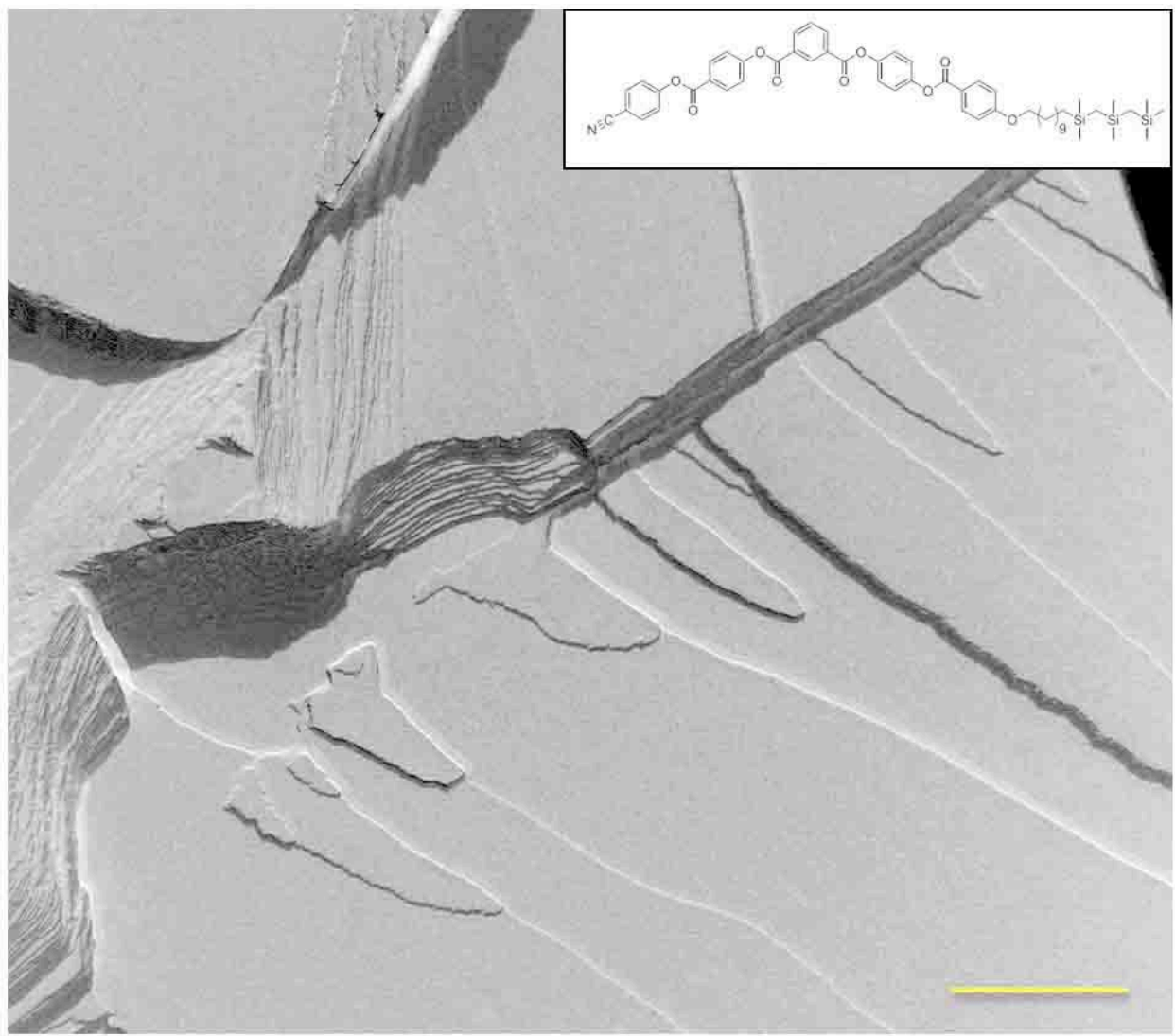

*Figure S2* - Fracture area of W623, shown, in its smectic A phase at T = 160°C. Here the layer planes are nearly parallel to the image plane as for the CB7CB layers in Figure S1. However, because of the density modulation and interlayer interfaces resulting from the the smectic layering the fractures follow the layer interfaces making the fracture surface smooth and terminated only with with distinct layer steps, in contrast to the rough surface observed in CB7CB in Figure S1. This difference can be taken as evidence for a layered structure without layer interfaces in CB7CB. Scale bar = 400nm.

*Figure S3*

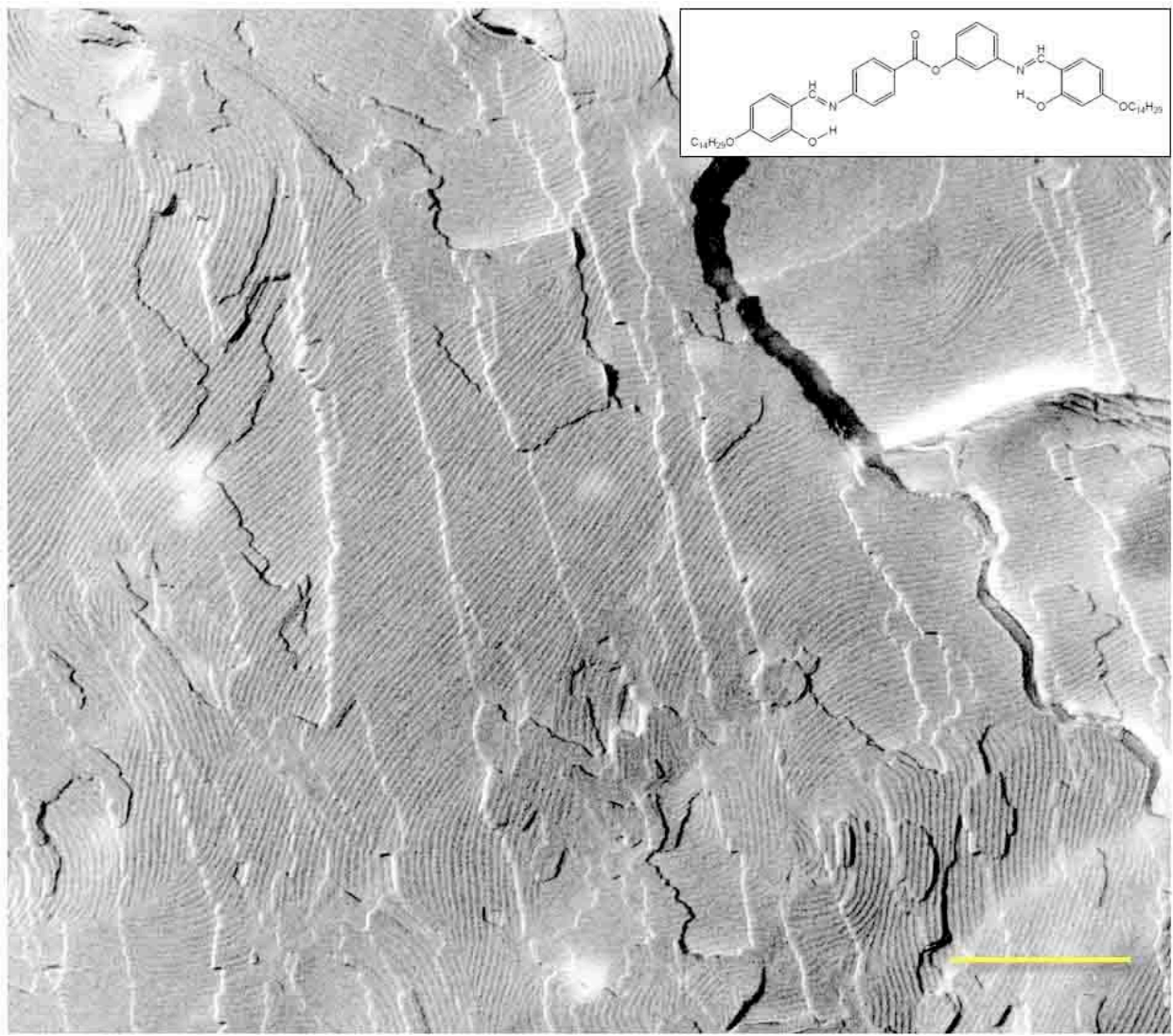

*Figure S3* - FFTEM image of the typical appearance of a 2D-ordered thermotropic columnar phase showing the 1D modulation of smooth layers terminated by layer steps. This image may be contrasted with that of CB7CB in Figure S1, which shows a much weaker tendency to form layer steps and smooth faces, which we take as evidence for a TB structure. Scale bar = 300nm.

*Figure S4*

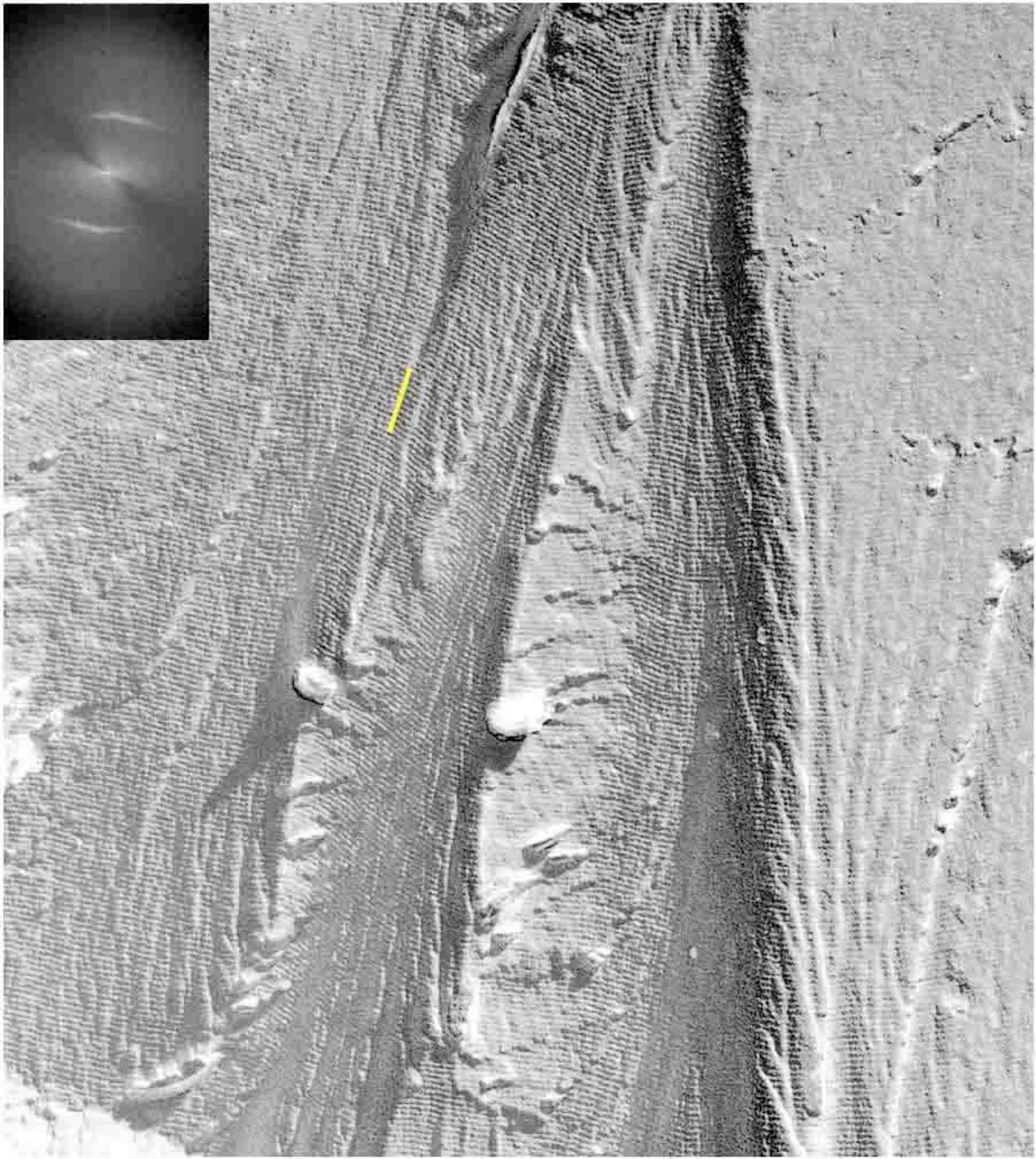

*Figure S4* - FFTEM image of CB7CB at T = 90ºC. Scale bar = 100nm.

*Figure S5*

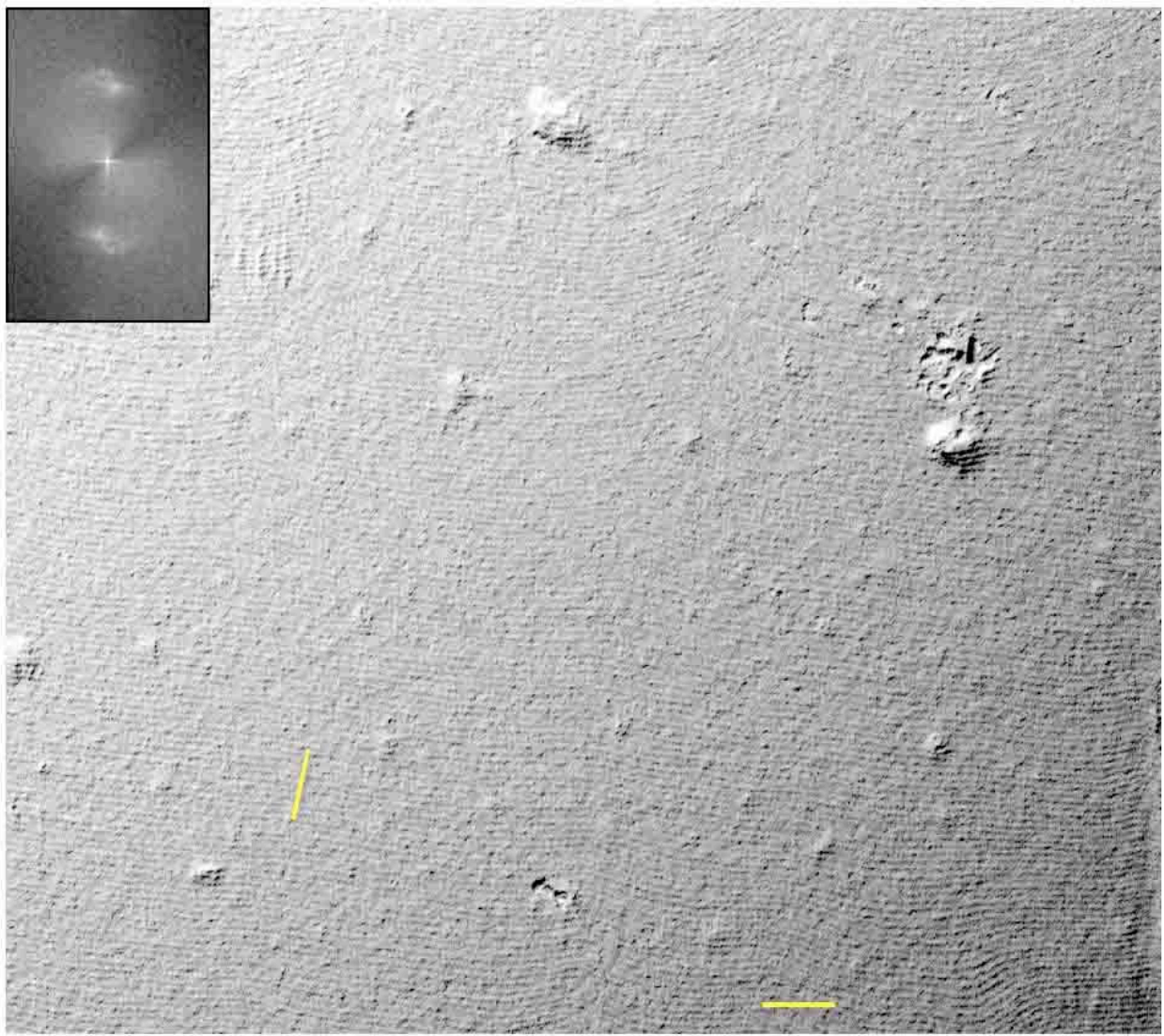

*Figure S5* - FFTEM image of CB7CB at T = 90°C. Here ψ, the angle between the layer normal and the fracture plane normal, is ψ ≈ 52°. Scale bar = 100nm.

*Figure S6*

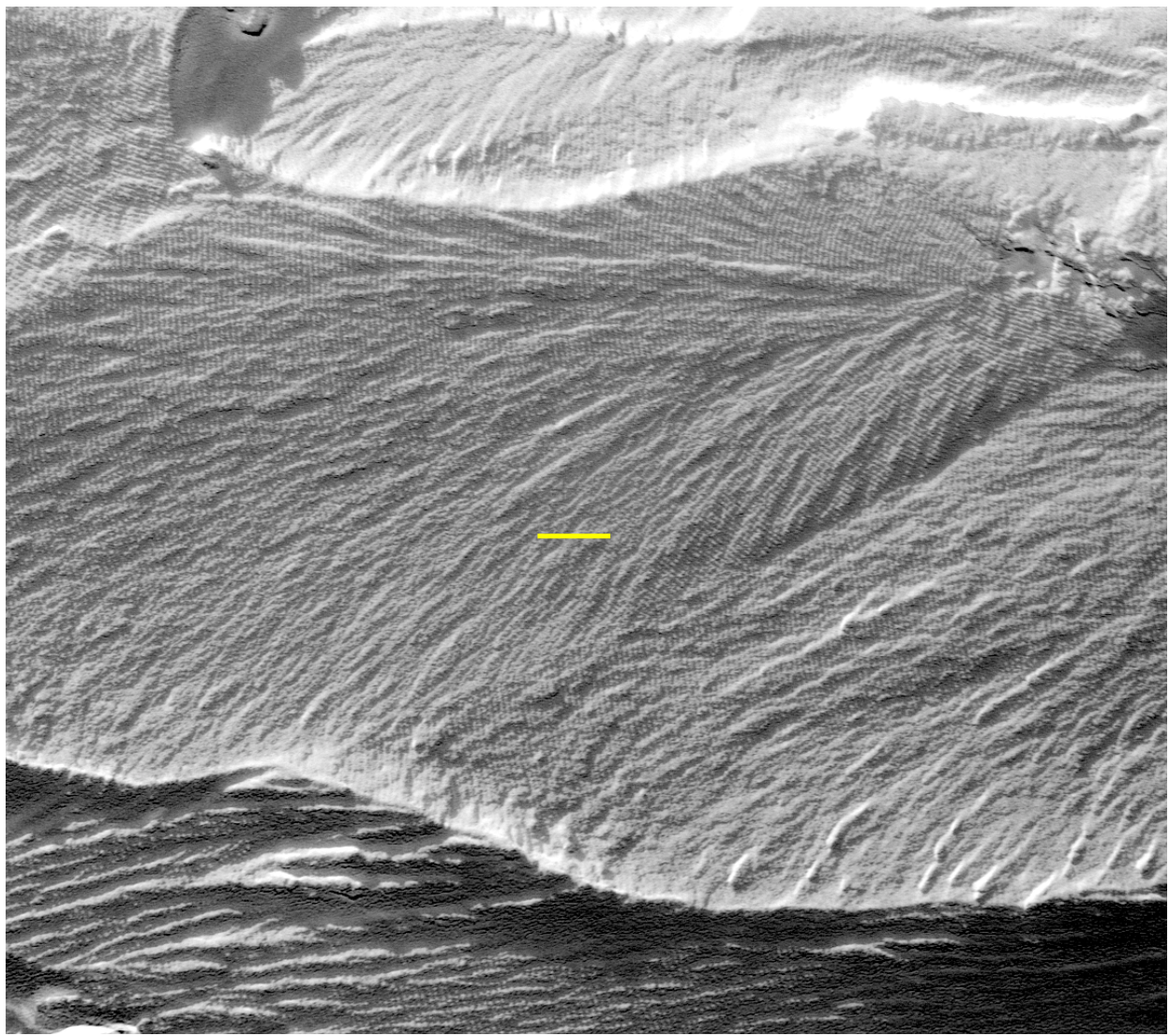

*Figure S6* - FFTEM image of CB7CB at T = 95°C. The general morphology of the modulation patterns and the period, $p_{TB}$, is similar to that at T = 90°C. Scale bar = 100nm.

*Figure S7*

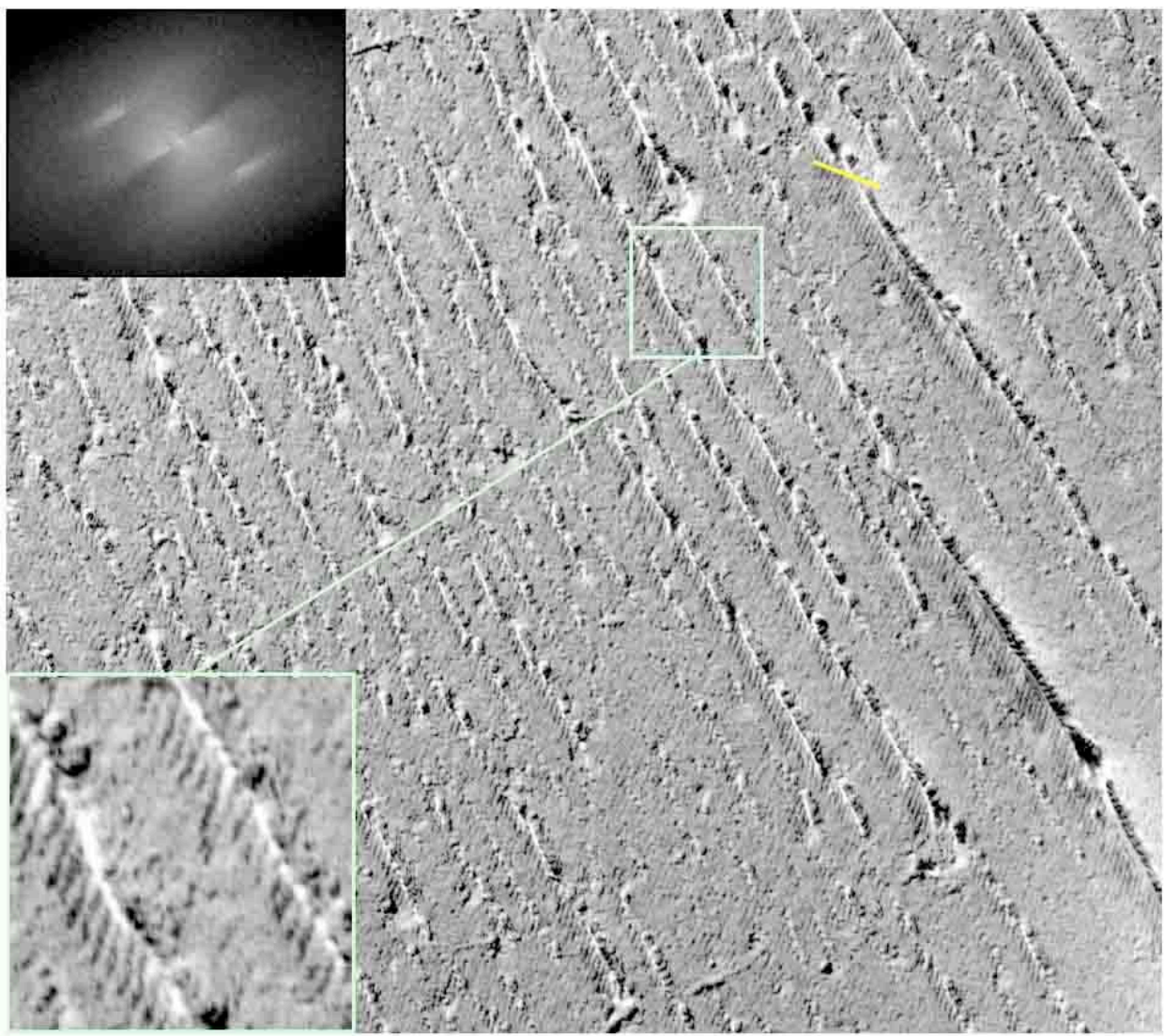

*Figure S7* - FFTEM image of CB7CB at T = 95°C exhibiting a chiral structure of the layer modulation and ridges in the fracture texture. This may be a structural manifestation of the inherent chirality of a TB phase, but any definitive interpretation of such an image will require a more detailed modeling of the fracture process. Scale bar = 100nm.

*Figure S8*

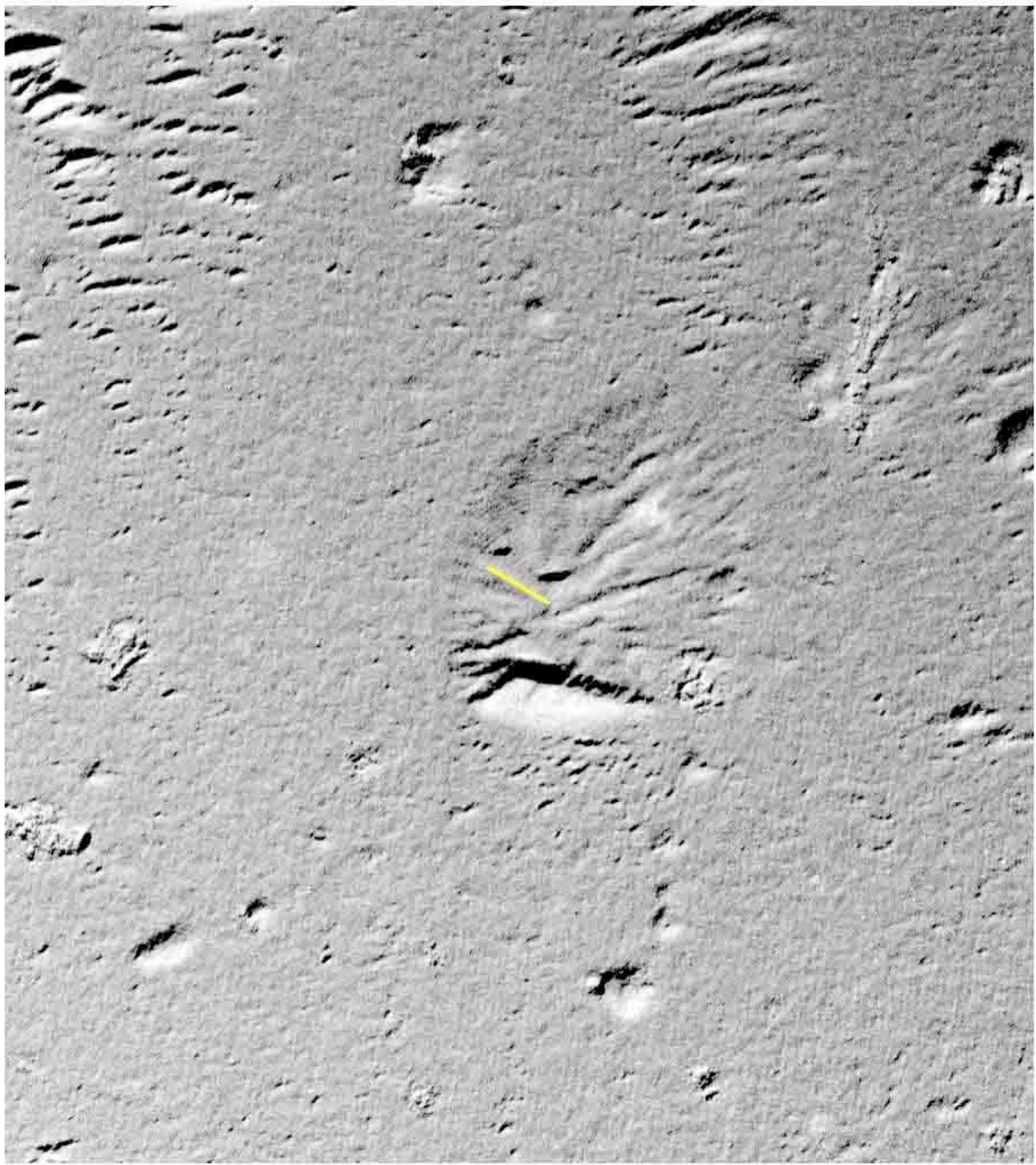

<u>*Figure S8*</u> - FFTEM image of CB7CB at T = 100°C, where large fracture face areas with stripes coexist with large areas without stripes, apparently because of coexistence of the NTB with the nematic. This is an area having stripes. The general morphology of the modulation patterns and the period, $p_{TB}$, is similar to that at T = 90°C. However, the layer modulation pattern on the fracture face is less distinct at this higher temperature near melting of the NTB to the nematic. Scale bar = 100nm.

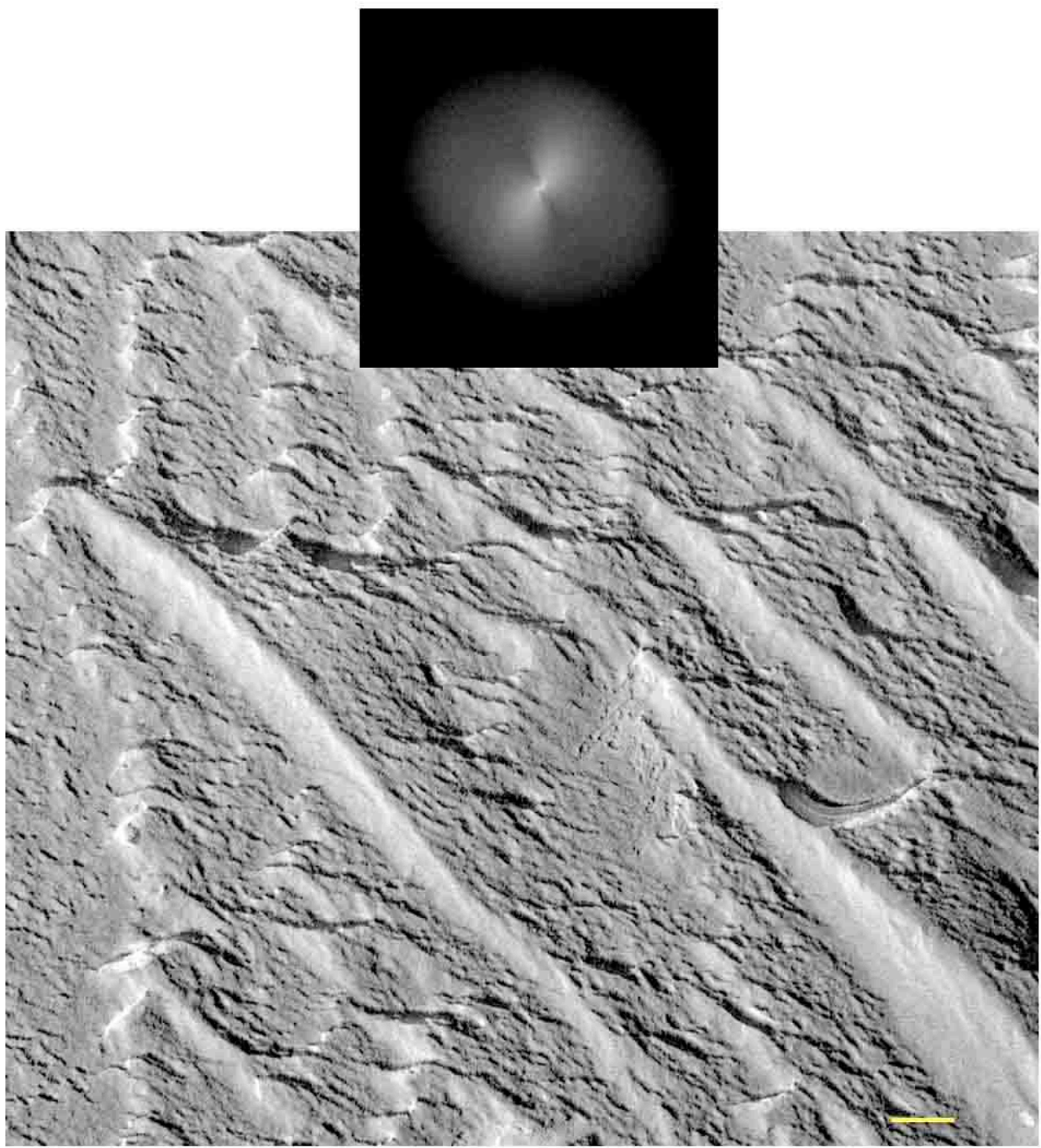

*Figure S9* - FFTEM image of CB7CB at T = 100ºC, where large fracture face areas with stripes coexist with large areas without stripes, apparently because of coexistence of the NTB with the nematic. This image and its Fourier transform show a domain without stripes. At T = 105ºC there are no remaining fracture face areas exhibiting stripes. Scale bar = 100nm.

*Figure S10*

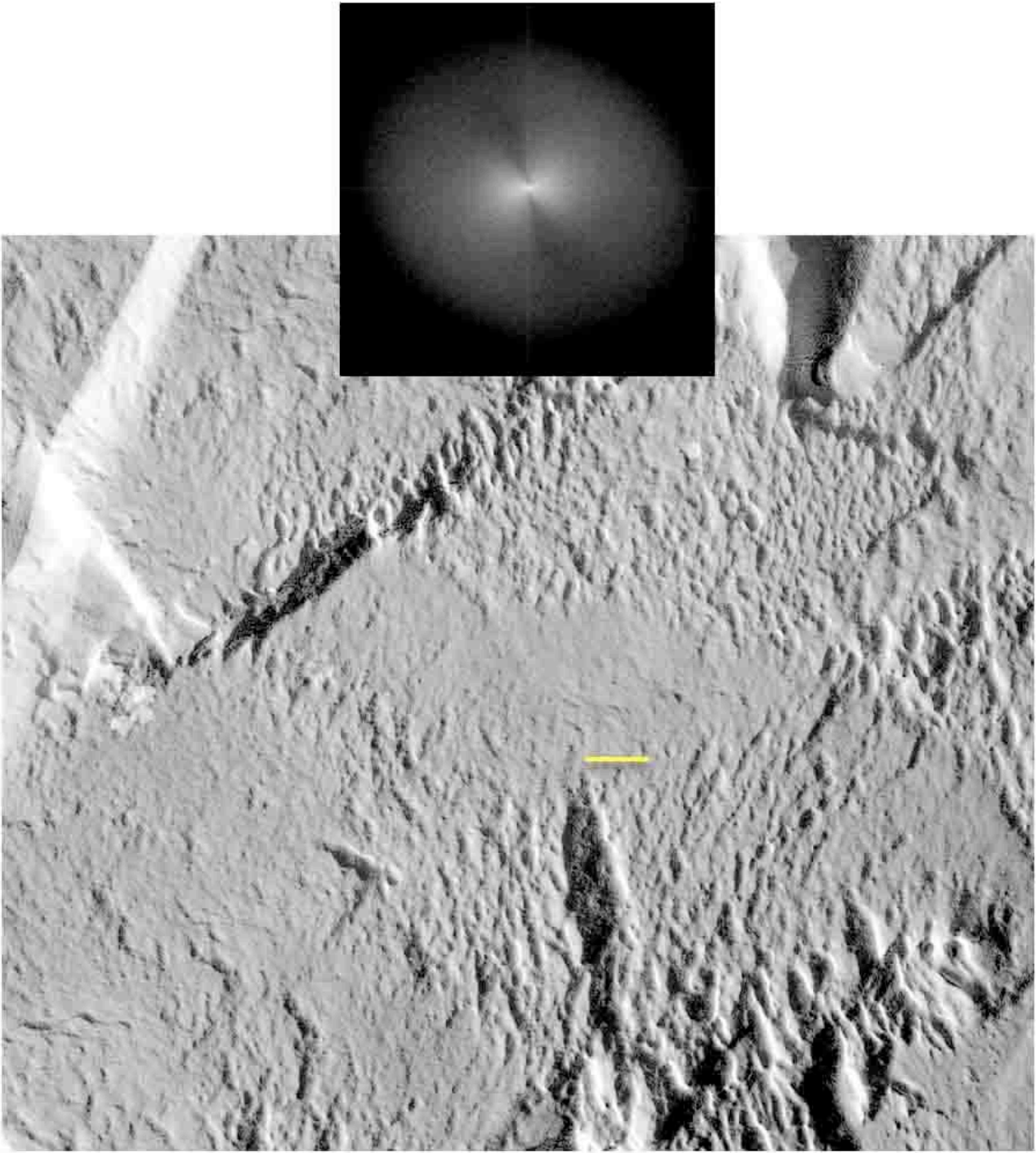

*Figure S10* - FFTEM image of CB7CB at T = 100ºC, where large fracture face areas with stripes coexist with large areas without stripes, apparently because of coexistence of the NTB with the nematic. This image and its Fourier transform show a domain without stripes. At T = 105ºC there are no remaining fracture face areas exhibiting stripes. Scale bar = 100nm.

*Figure S11*

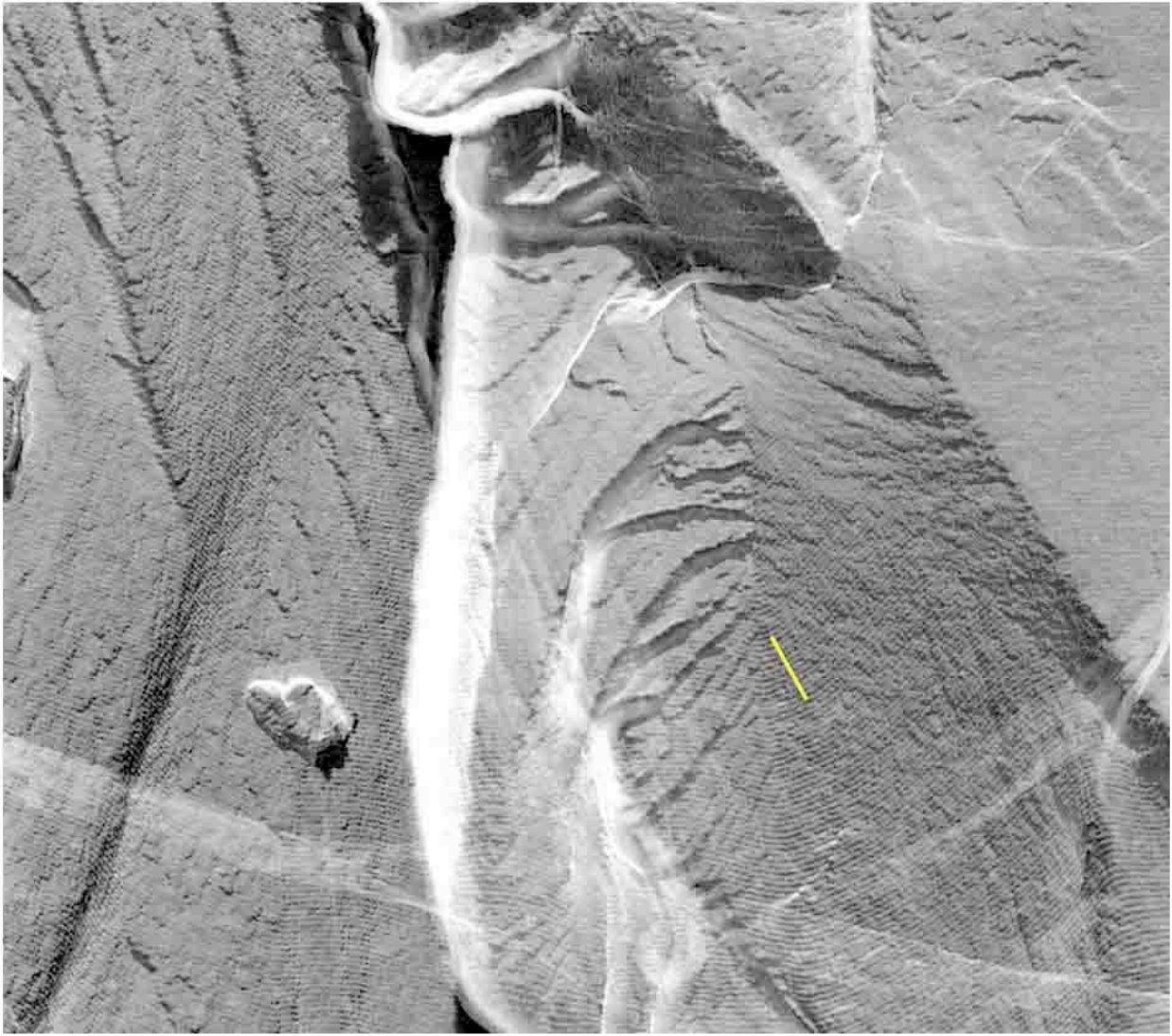

*Figure S11* - FFTEM image of CB7CB at T = 29°C, in the supercooled NTB phase. The general morphology of the modulation patterns is similar to that at T = 90°C. The period, $p_{TB}$, is slightly larger. The layer modulation pattern on the fracture face becomes more distinct with decreasing temperature and the fracture surfaces where the layers are normal to the fracture plane become smoother than at T = 90 °C. This suggests that preference for fracture normal to the layers is stronger in the NTB glass. Scale bar = 100nm.

*Figure S12*

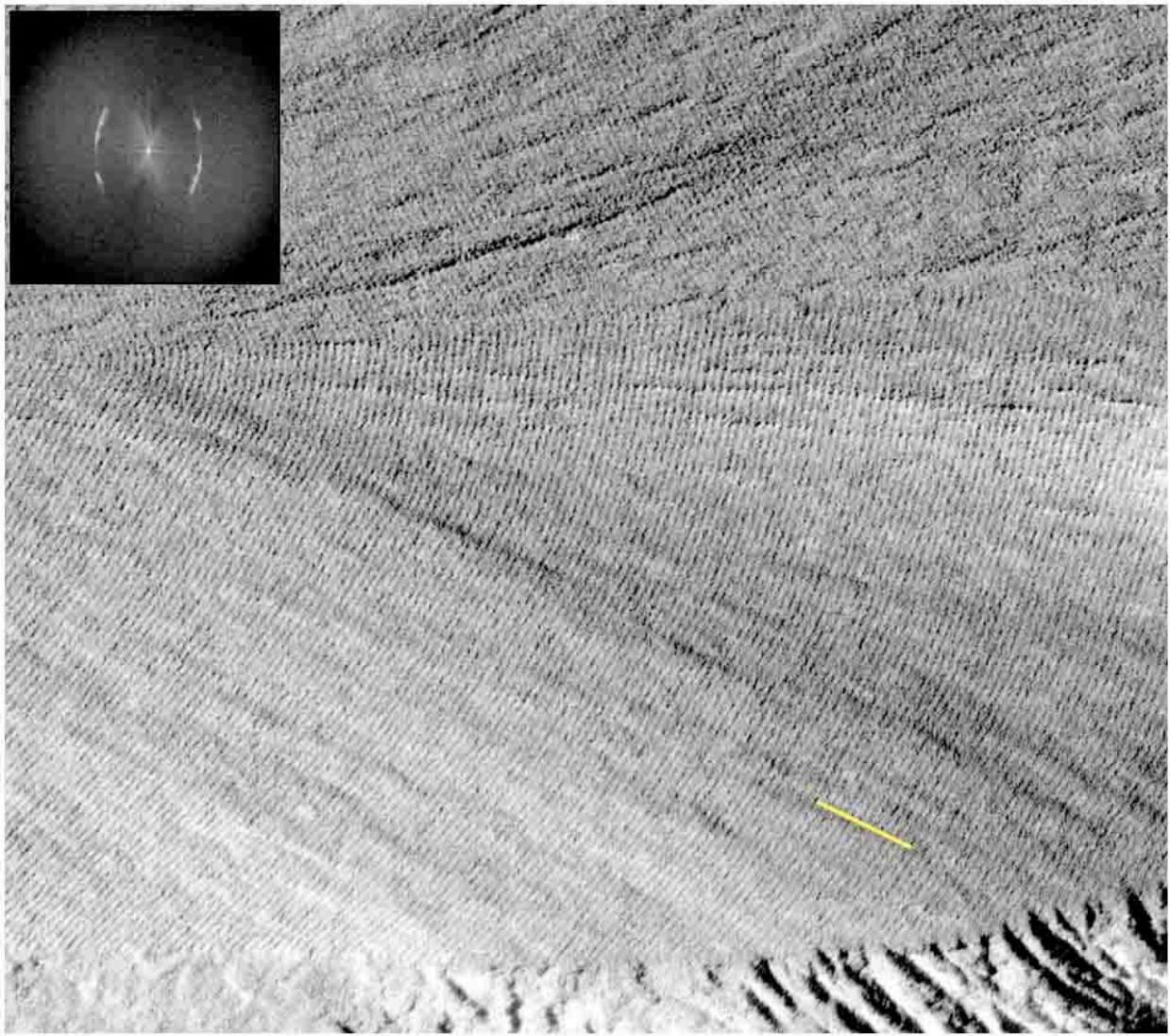

*Figure S12* - FFTEM image of CB7CB at T = 29°C, in the supercooled NTB phase. Scale bar = 100nm.

*Figure S13*

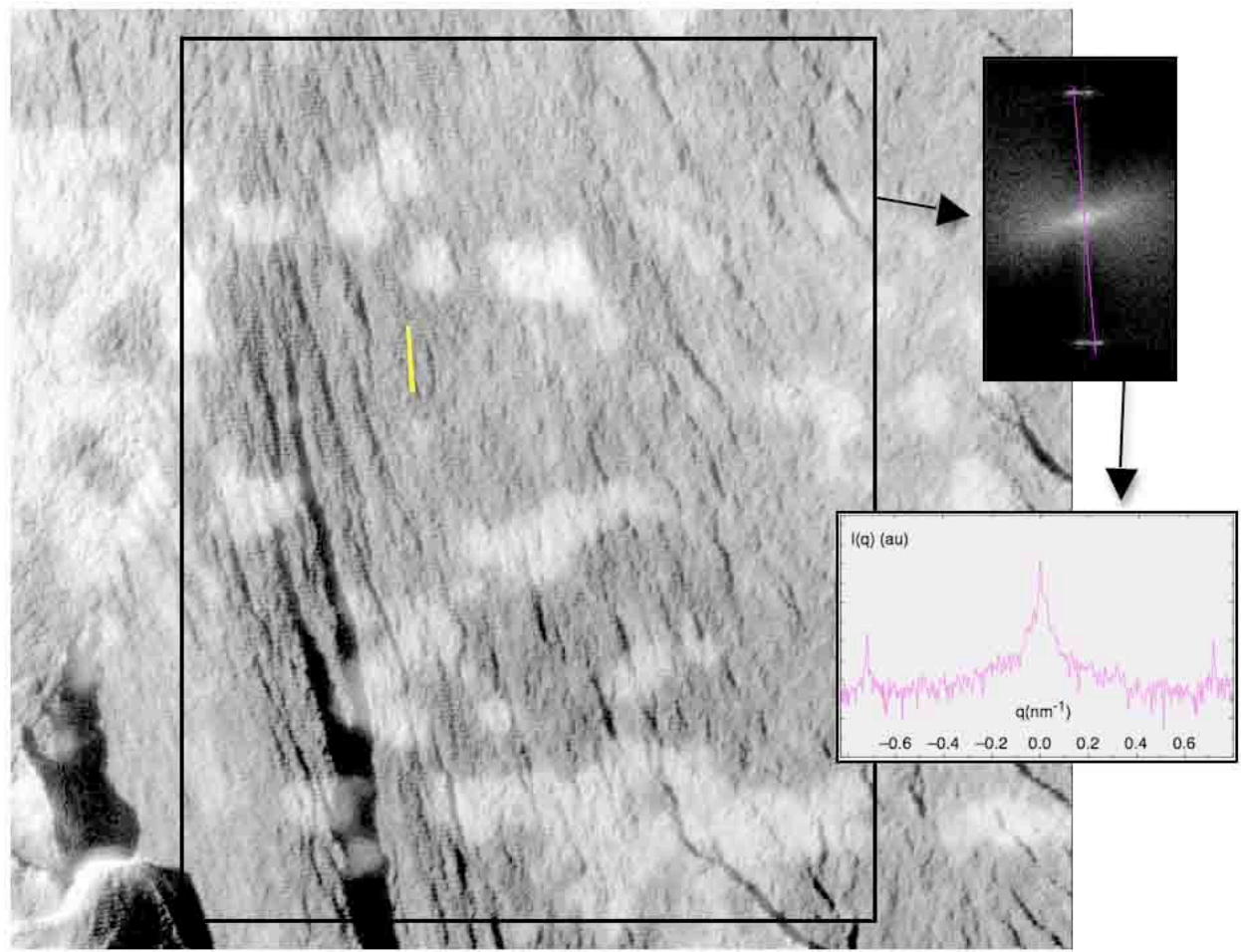

*Figure S13* - FFTEM image of CB7CB at T = 29°C, in the supercooled NTB phase, showing the Fourier transform of the intensity distribution of the image in the selected (black box) area, and the profile of the Fourier transform along the magenta line, showing the first order Bragg reflection peaks from the layer modulation. The modulation peaks are sharp, of HWHH ~ 1 pixel in q-space, indicating that the coherence length for loss of layer periodicity in the direction normal to the layers is longer than the FFT resolution. Scale bar = 100nm.

*Figure S14*

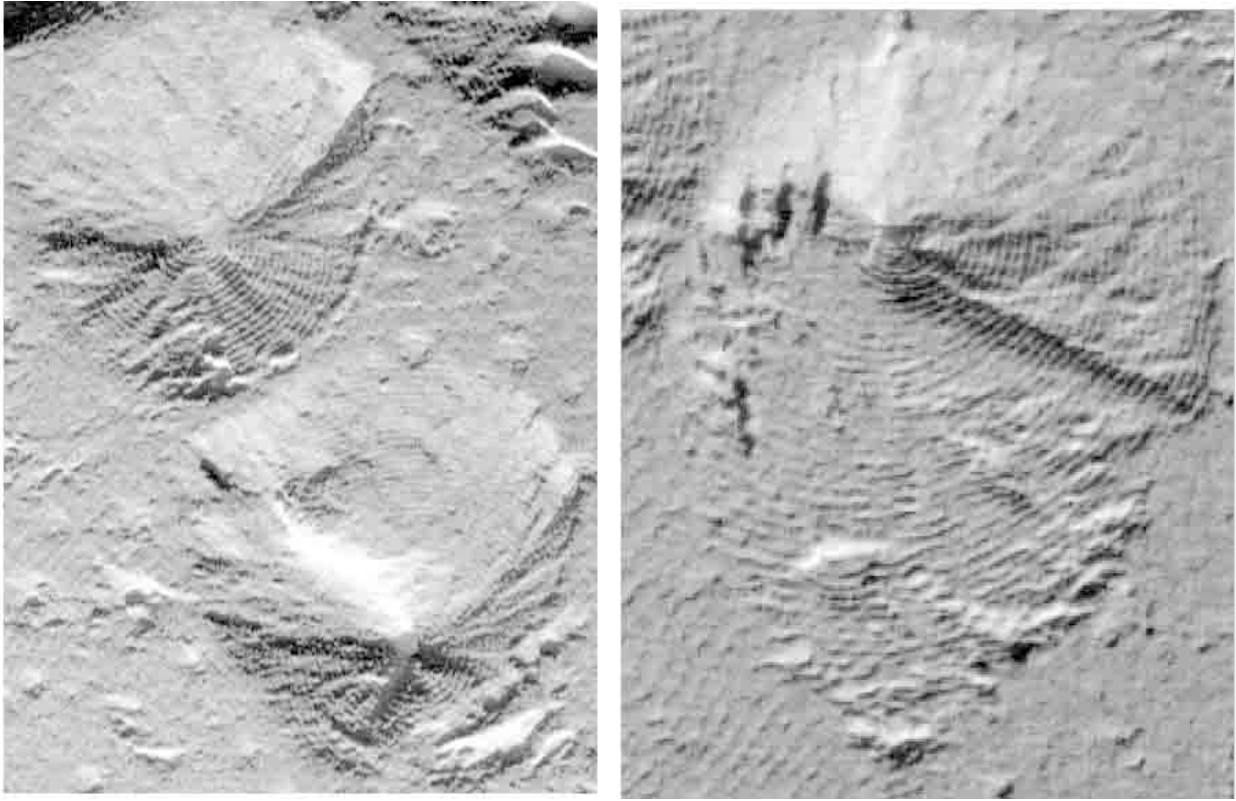

*Figure S14* - FFTEM image of CB7CB NTB phase, showing commonly observed circular focal conic domains that produce a cone shaped hill where they terminate in the fracture plane (left). Some of these have a distinctly teardrop shape (right).

-19-

*Figure S15*

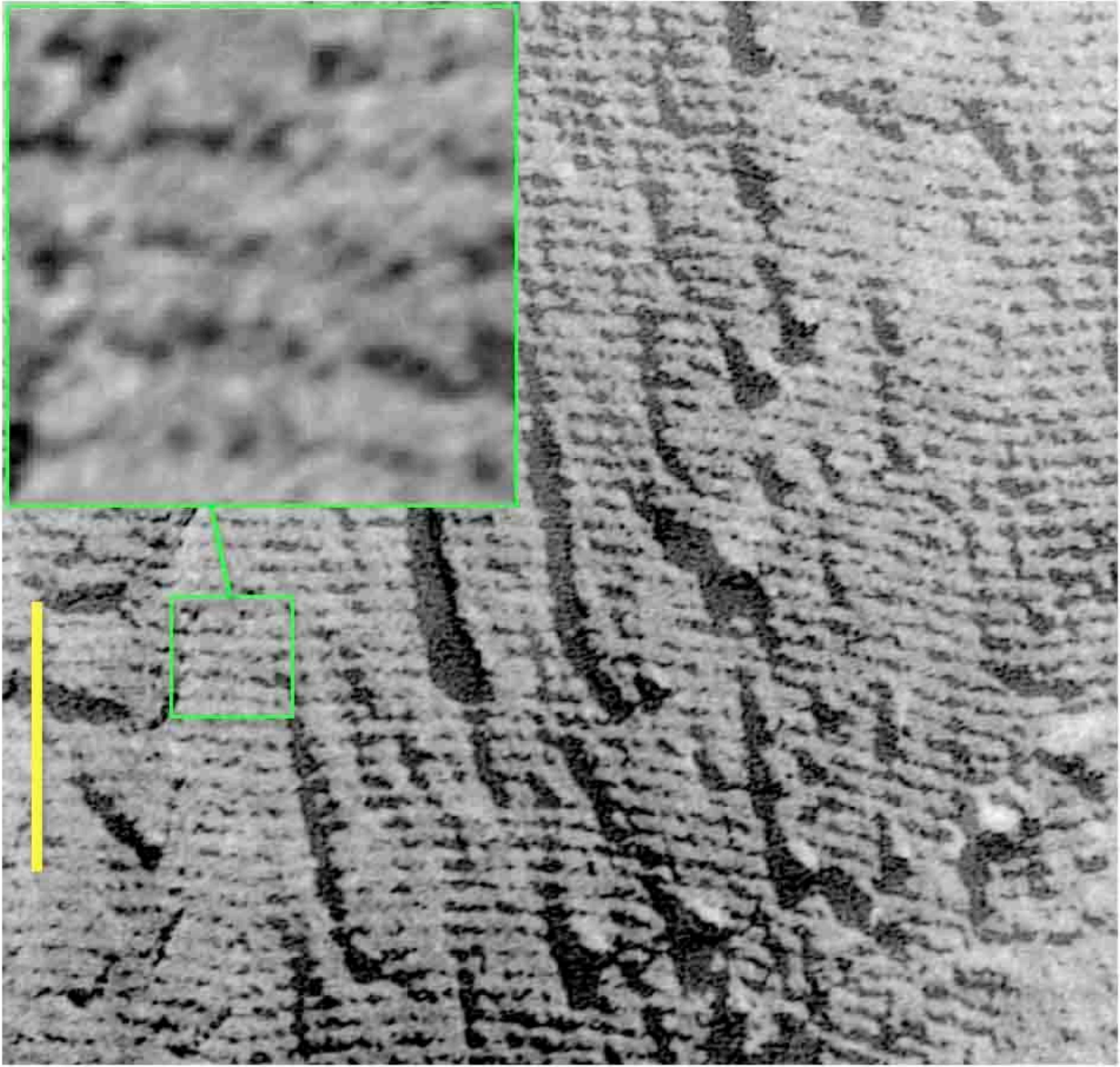

*Figure S15* - Additional views of the higher magnification image shown in Figure 4C. The inset shows the oblique striations giving the oblique diffuse oval hump around q = 0 in the FFT of Figure 4C. These structures may reflect the local chirality of the TB helix, but confirmation of this will require higher resolution techniques. The inset also serves to indicate the inherent resolution of the freeze fracture technique, here on the order of 0.5nm. Scale bar = 100 nm.

*Figure S16*

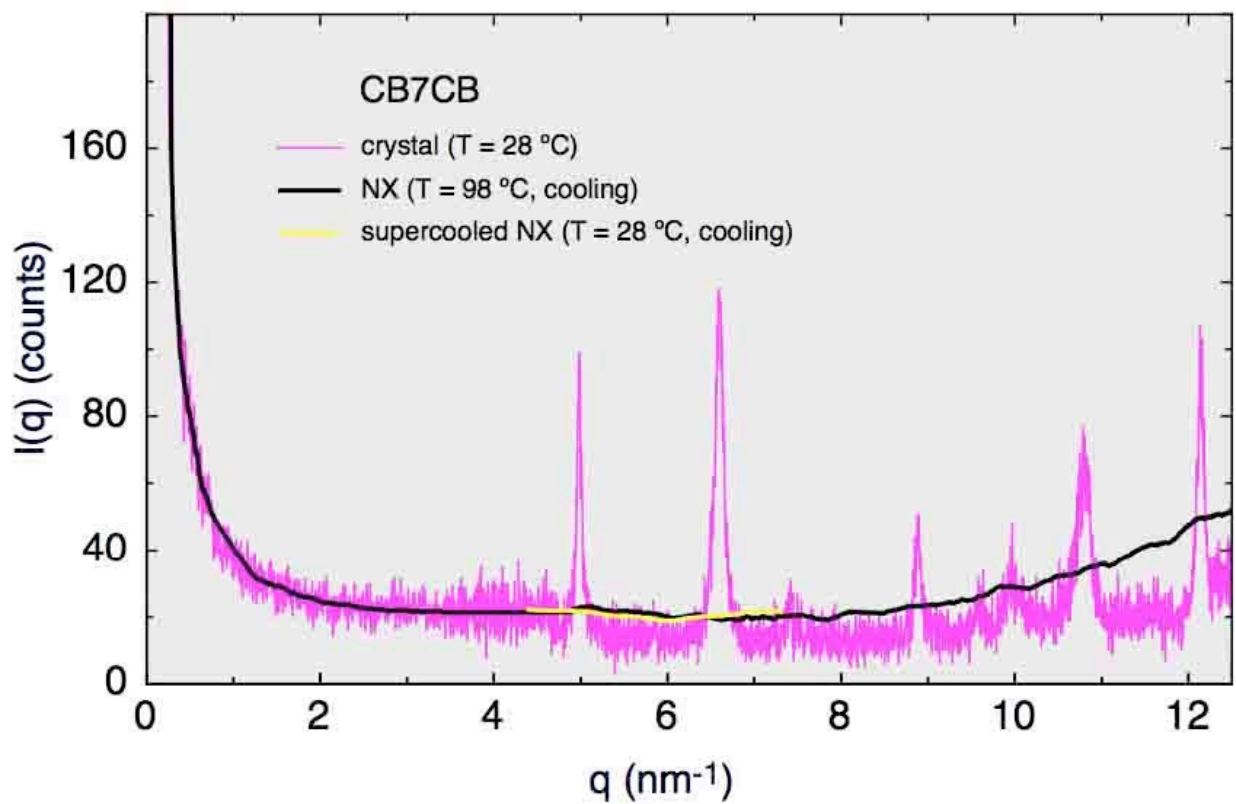

*Figure S16* – Powder XRD scans of the crystal phase of CB7CB at room temperature (magenta), the NTB phase, obtained upon cooling from the isotropic and nematic (black), and the supercooled NTB phase (yellow). The scan in the supercooled NTB phase, over the limited range, $0.45\,\text{nm}^{-1} < q < 0.7\,\text{nm}^{-1}$, probes the intensity of the two large crystal peaks. Upon cooling from the NTB to room temperature at an intermediate rate, i.e., over a 1 to 60 second interval, the thermotropic NTB phase supercoools into an NTB glass, as evidenced by the complete absence of the crystal peaks. These may reappear after several days, or as diffuse reflections for very slow cooling.